\documentclass[journal]{IEEEtran}
\usepackage{cite}
\usepackage{graphicx}
\usepackage{tikz}
\usepackage{xcolor,colortbl}
\usepackage{amsmath}
\usepackage{multirow}% http://ctan.org/pkg/multirow
\usepackage{hhline}% http://ctan.org/pkg/hhline
\usepackage{color}
\usepackage[flushleft]{threeparttable} % http://ctan.org/pkg/threeparttable
\usepackage{booktabs,caption}
\usepackage{tablefootnote}
\ifCLASSINFOpdf
 \else
\fi
\begin{document}
\title{PinMe: Tracking a Smartphone User around the World}
\author{Arsalan Mosenia, \emph{Student Member, IEEE,} Xiaoliang Dai, Prateek Mittal, 
\emph{Member, IEEE,} and Niraj K. Jha, \emph{Fellow, IEEE}
\thanks{ 
Arsalan Mosenia, Xiaoliang Dai, Prateek Mittal, and Niraj K. Jha are with the Department of Electrical Engineering, Princeton University, Princeton, NJ 08544, USA (e-mail: {arsalan$||$xdai$||$pmittal$||$jha}@princeton.edu).
}

}

\maketitle

\begin{abstract}
With the pervasive use of smartphones that sense, collect, and process 
valuable information about the environment, ensuring location privacy has become one of the most important concerns in the modern age.

A few recent research studies discuss the feasibility of processing sensory 
data gathered by a smartphone to locate the phone's owner, even when the user 
does not intend to share his location information, e.g., when the user has 
turned off the Global Positioning System (GPS) on the device. Previous 
research efforts rely on at least one of the two following fundamental 
requirements, which impose significant limitations on the adversary: (i) the 
attacker must accurately know either the user's initial location or the set of 
routes through which the user travels and/or (ii) the attacker must measure a 
set of features, e.g., device acceleration, for different potential 
routes in advance and construct a training dataset. 

In this paper, we demonstrate that neither of the above-mentioned requirements 
is essential for compromising the user's location privacy. We describe PinMe, a 
novel user-location mechanism that exploits non-sensory/sensory data stored on 
the smartphone, e.g., the environment's air pressure and device's timezone,
along with publicly-available auxiliary information, e.g., elevation maps,
to estimate the user's location when all location services, e.g., GPS, are 
turned off. Unlike previously-proposed attacks, PinMe neither requires any 
prior knowledge about the user nor a training dataset on specific routes. We 
demonstrate that PinMe can accurately estimate the user's location during four 
activities (walking, traveling on a train, driving, and traveling on 
a plane). We also suggest several defenses against the proposed 
attack.
\end{abstract}


\IEEEpeerreviewmaketitle

\section{Introduction}
\label{INT}
With widespread use of smartphones that can sense and collect environment-related 
data and process them to extract valuable information about the environment, 
ensuring privacy has become one of the most important challenges in 
the modern era. Indeed, rapid technological advances in electronics and mobile 
devices have led (and will continue to lead) to serious concerns about privacy 
in general, and location privacy in particular \cite{SPAM1}.

Modern smartphones are equipped with many compact sensors, e.g., 
accelerometers and barometers, and powerful communication capabilities in order 
to offer a variety of services. Although the numerous smartphone 
applications make the user's life convenient, they can also 
intentionally/unintentionally reveal personal or corporate 
secrets \cite{SEC1,SEC2,SEC3,SEC4,SEC5,SEC6,SEC7,NAVEED}. In particular, they 
can leak valuable data about the user's whereabouts, which can be 
processed to extract contextual information about his habits, regular 
activities, and even relationships \cite{LSEC1,LSEC2}. Moreover, disclosure 
of the user's location may expose him to location-based spams, scams, and 
advertisements, or make him a victim of blackmail or 
violence \cite{SPAM1,SPAM2}.

With the emergence of enormous privacy concerns in the last decade, several 
privacy policies have been put in place to force organizations to take their 
users' privacy into account. In particular, the U.S. Congress introduced the 
Geolocation Privacy and Surveillance Act in 2011 to 
provide a legal framework for giving government agencies, commercial entities, 
and private citizens clear guidelines for when and how geolocation information 
can be accessed and used \cite{ACT}. As a result, in all modern smartphones, 
an application must explicitly ask for the user's permission if it wants to 
access location services, e.g., GPS \cite{LOC1,LOC2}.

A few recent research efforts have demonstrated the 
feasibility of locating smartphone owners without accessing 
GPS \cite{ATT1,BON,SUB,NAVEED}. For instance, Michalevsky et al. proposed 
PowerSpy \cite{BON}, a mechanism that locates the user by processing the power 
consumption of the smartphone, when the user travels through a 
known set of routes. PowerSpy was able to detect $45\%$ of driving trajectories
in the best-case scenario. Han et al. showed that accelerometer 
readings can be used to estimate the trajectory and starting point of an 
individual who is driving \cite{ACCO}. They were able to return two clusters of possible starting points (each including five points) such that the starting point was within one of the clusters.

The successful demonstration of such attacks against location privacy suggests that revealing the user's location by processing presumably non-critical data is feasible. However, all previously-suggested attacks against location privacy mainly rely on at least one of the three following fundamental requirements. 

\begin{itemize}
\item  \textbf{Req. 1:} The attacker must either know the user's initial location (the exact GPS coordinates) or has \textit{substantial prior knowledge} of the area through which the victim is traveling, e.g., the attacker assumes that the victim is traveling through a small set of known routes.

\item \textbf{Req. 2:} The attacker must measure a set of features, e.g., power consumption \cite{BON}, for different \textit{potential routes} in advance and construct an attack-specific training dataset.

\item \textbf{Req. 3:} The sensory data must be continuously collected at 
a \textit{high sampling rate}, e.g., $30Hz$ \cite{ACCO}, \cite{BARO}, which 
is significantly higher than the sampling rate needed for a majority of benign 
applications. 

\end{itemize}

The first two requirements significantly limit the attacker's ability to 
locate the user in realistic scenarios, and the third can raise suspicion, 
making it easier to detect the attack \cite{LEAVE_ME}. Even with these 
requirements, previous attacks offer a rough estimation of the user's 
trajectory, as discussed later in Section \ref{RELATED}.

This paper aims to demonstrate that none of the above-mentioned requirements 
is needed to accurately track the user when all location services, e.g., GPS, 
are off. We propose an attack on location privacy in which: (i) the attacker 
needs neither the user's initial location nor a small set of potential travel 
routes, (ii) he is not burdened with the construction of an attack-specific 
database, and (iii) he does not collect data at a high sampling rate,
e.g., as demonstrated later, a sampling rate of $0.1Hz$ is sufficient to track 
the user when he is driving. The first two characteristics of the proposed 
scheme enable an attacker to launch an attack on a large scale, when he has 
no prior knowledge about users' initial locations or the set of routes through 
which he travels. The third one makes the attack invisible to known defenses 
that detect the maliciousness of an application based on its high sampling 
frequency, e.g., the defense in \cite{LEAVE_ME}. 

\begin{figure}[t]
\centering
\includegraphics[trim = 0mm 10mm 0mm 10mm ,clip, width=240pt,height=180pt]{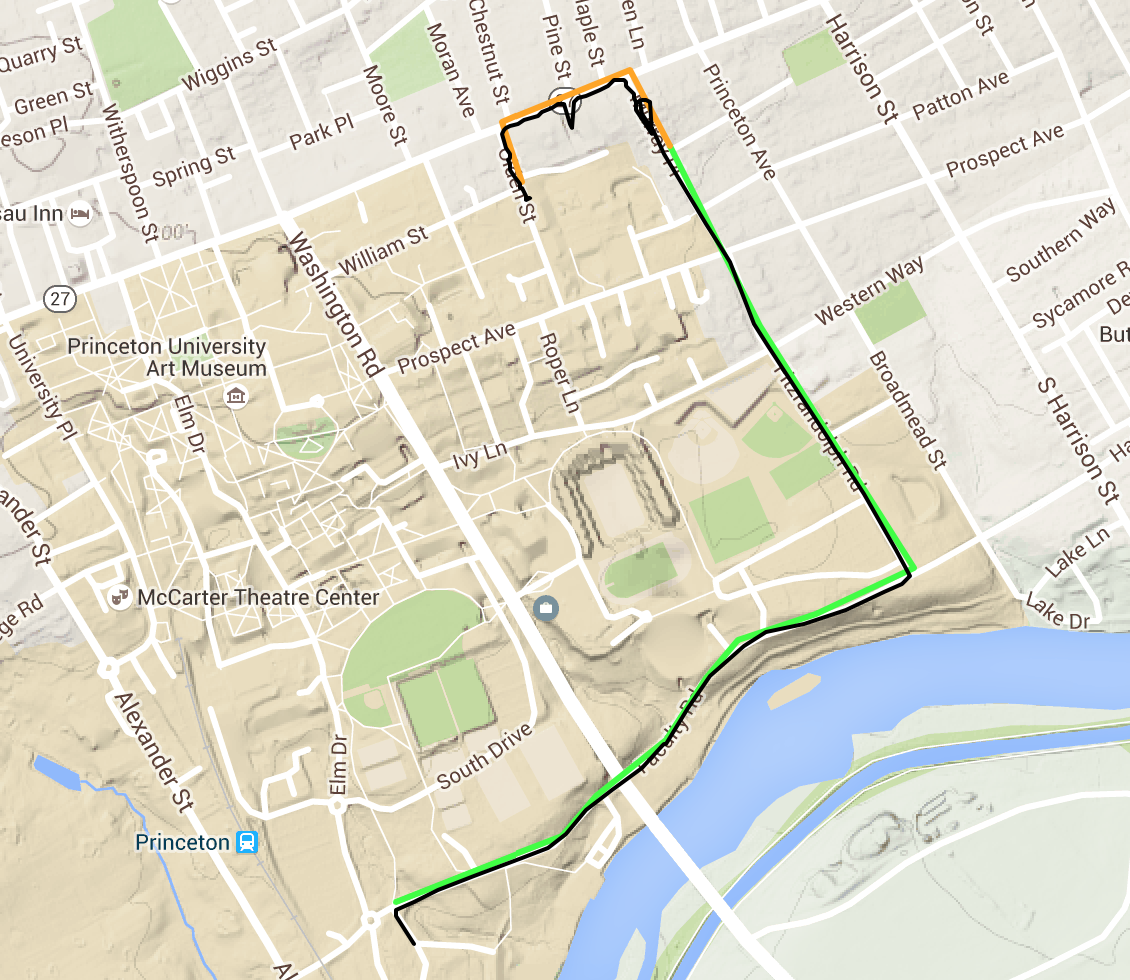}
\caption{PinMe could find and return the user's trajectory without accessing 
GPS data. The green and orange lines demonstrate the estimated paths
traversed by the user during driving and walking, respectively. The black line 
is the actual user trajectory reported by GPS data.} 
\label{fig:traGPS}
\end{figure}

Our main contributions can be summarized as follows:
\begin{enumerate}
\item We develop PinMe, a location mechanism that enables an attacker to accurately locate the user using sensory/non-sensory data along with publicly-available auxiliary information. 
\item We demonstrate how different types of seemingly-benign non-sensory
data, e.g., the smartphone's timezone and network status, and sensory
data, e.g., air pressure and heading, can offer sensitive information to 
the attacker who aims to locate the user. 
%To the best of our knowledge, %PinMe is the first location mechanism that uses air pressure readings for %outdoor positioning.
\item We introduce five sources of publicly-available auxiliary information (public maps, transportation time tables, airports' specification databases, weather reports, and trains' heading dataset) that can be used in conjunction with smartphone's data to develop an attack against location privacy. 
\item Unlike previously-proposed attacks \cite{BON,ACCO}
that are focused on a single activity, e.g., driving, we demonstrate how a 
user can be located when he is: (i) traveling on a plane, (ii) walking, 
(iii) traveling on a train, and (iv) driving. As far as we know, PinMe is the 
first smartphone-based user location mechanism that aims to locate the
user while undertaking different activities.  
\item In order to evaluate the accuracy of the proposed location mechanism, 
we collect real-world data using three devices (iPhone 6, iPhone 6S, and 
Galaxy S4 i9500).

\item  We evaluate the accuracy of PinMe for estimating the user's location 
using two real-world datasets. We demonstrate that, unlike previous attacks, 
PinMe is able to accurately and uniquely return a trajectory that is 
comparable to GPS-based trajectory (Fig.~\ref{fig:traGPS}).
\item Finally, we discuss defenses against the proposed attack.
\end{enumerate}

\textit{To sum up, PinMe aims to offer a comprehensive (i.e., covering
multiple activities) attack that minimizes the need to have prior knowledge 
about the user, removes the need for building attack-specific datasets, and 
uses the interdependence between seemingly-independent activities to obtain
an accurate user trajectory. Our end-to-end evaluation demonstrates that 
PinMe works accurately (comparable to GPS) in real-world scenarios. As 
discussed in Section \ref{CONT}, protecting the user against this attack can 
be very challenging due to its robustness against potential sources of noise 
and the low sampling rate required for the attack.}

The remainder of this paper is organized as follows. Section \ref{THR} 
provides the problem definition and discusses how the attacker can acquire 
the data needed for the proposed attack. Section \ref{PMECH} discusses PinMe 
comprehensively and describes different sources of information and algorithms 
used to implement the attack. Section \ref{EVAL} describes how we collected 
real-world data for evaluating PinMe and examines the accuracy of the proposed 
location mechanism. Section \ref{CONT} suggests several countermeasures for 
mitigating the risks of the proposed attack. Section \ref{RELATED} summarizes 
related work. Section \ref{DISC} discusses the limitations of PinMe and 
describes how we used the interdependence between activities to facilitate (and enhance the accuracy of) our proposed attack and how PinMe can be used as an alternative to GPS in autonomous cars to 
enhance their security. Finally, Section \ref{CONC} concludes the paper.

\section{Threat Model}
\label{THR}
In this section, we first describe several consequences of launching an attack 
against location privacy, and provide a brief description of our proposed 
attack. Then, we discuss how attackers can acquire the data that are 
required to launch the proposed attack. 

\subsection{Problem definition}
Today's smartphones are equipped with several low-power high-precision sensors 
and powerful processors that enable them to continuously collect and process 
environment-related data. As a result, a modern smartphone carries several 
types of valuable data. Such data can be processed to reveal sensitive 
information about the 
phone's user. For example, the contextual information attached to movement 
traces conveys much about the user's interests, activities, and even 
relationships \cite{SPAM1}. 

Launching an attack against location privacy can expose the user to unwanted 
advertisement, spams, or scams. Moreover, it can lead to several consequences, 
ranging from the uncomfortable feeling of being monitored to unwanted 
disclosure of personal activities or even actual physical harm
\cite{PHYDAM}. For example, it may be embarrassing for a user if his/her 
relatives find out that he/she went to certain places, e.g., an HIV clinic
or an abortion clinic. While these consequences are a direct result of 
manual inspection of leaked location-related information, several recent 
research efforts have investigated the feasibility of extracting 
other valuable information from the user's location-related information. For example, early research work in this area \cite{HOH} explored the possibility 
of inferring information about the user's habits and detecting places important to him, e.g., his home and office, from GPS traces.

Although the importance of preventing location services, e.g., GPS, from 
leaking unwanted information has become clear, the extent of location-related information that can be inferred from presumably non-critical data, such as movement-related data, e.g., acceleration and heading, and environment-related data, e.g., air pressure, is neither well-known nor well-understood. This paper aims to demonstrate the possibility of accurately locating the smartphone's user using such presumably non-critical data stored on the phone. 

\subsection{Acquiring data}
\label{AQU}
The attacker can obtain the smartphone's non-sensory and sensory data, which 
are required for the proposed attack, using one of the two following approaches:\\ 
\noindent \textbf{Approach 1: Utilizing a malicious application}\\ 
\indent Smartphones are characterized by their ability to run third-party 
applications. Both Android and iOS offer hundreds of thousands of applications 
through their application markets. Such markets benefit developers by 
simplifying application sales and distribution. The existence of huge 
application markets might also enable cyber criminals to distribute a 
malicious application in an attempt to steal personal information stored on 
the phone, e.g., credit card numbers and personal photos. Fortunately, such 
critical information is commonly protected by the smartphone's operating 
system, and users are also very careful about sharing their personal 
information with third parties. However, several types of non-sensory/sensory 
data, which are stored on the smartphone, are either loosely-protected or not 
protected at all, e.g., gyroscope, accelerometer, barometer, and magnetometer 
measurements are accessible by an application installed on the smartphone 
without requiring user's approval. As a result, a malicious application that is installed on the smartphone and runs in the background can continuously capture such data without arousing suspicion. \\

%Such an application can either transmit raw 
%non-sensory/sensory data to the attacker's server or perform some computation 
%on the data (e.g., compression) and only transmit the results of the 
%computation. \\ 
\noindent \textbf{Approach 2: Accessing a presumably-trusted application 
server}\\
\indent Several trusted applications upload their data to the cloud. For 
example, the majority of fitness monitoring applications continuously collect 
and upload the user's data to the cloud.  The collection of the data in the 
cloud enables the user to  access and share his fitness statistics with his 
family, friends, and peer groups. A recent report by the mHealth development 
industry \cite{REPORT} estimates that there are about 100,000 applications 
dedicated to health and fitness. Such applications can, without arousing 
suspicion, collect and upload a significant amount of valuable 
non-sensory/sensory data, which can be post-processed to infer critical 
information about the user. In particular, as we demonstrate later, 
an attacker, who can access such application databases, e.g., the application 
development company or an individual who has access to the data shared by the 
user, can post-process such data to estimate the past locations of the user.

\textbf{Our approach:} In this paper, we assume that the proposed location 
mechanism obtains the required non-sensory/sensory data using the first 
approach. In fact, we installed an application on the smartphone that 
continuously collects the required data. We assume that the application does 
not have access to GPS. Moreover, the application has no permission to query 
the identity of visible cellular base stations or the service set identifier 
(SSID) of visible WiFi networks. To sum up, we assume that the attacker only 
uses presumably \textit{non-critical data collected by a malicious application 
along with publicly-available auxiliary information to reveal the user's 
location.} 
The proposed attack does not rely on careless behaviors of the user (e.g., a careless user may just accept all permission requests, including a request to access GPS data, without carefully reviewing them). In fact, PinMe aims to demonstrate the feasibility of a privacy attack against careful users (for example, a user who checks what he shares with third-party applications, minimizes the access level of untrusted applications, and even turns off all location services when he travels through sensitive routes to ensure his location privacy). The introduction of this attack sheds light on the possibility that a third party, which 
has an application on the user's smartphone, can potentially extract his 
sensitive location information without asking for any permission (except 
Internet connectivity that is needed for sending either raw data or inferred 
location to the third party). 

%The cloud has become as ubiquitous as a basic utility, as data can be accessed anywhere in the world with an Internet connection. 

%We have processed the data on a laptop computer. However, the
%processing algorithms discussed in this paper are light-weight and
%could be also implemented on the smartphone. In that situation the
%malicious application only needs to transfer a small amount of
%information to the attacker, e.g., the time of activities and 
%starting/ending positions.

\begin{table*}[t]
\caption{Smartphone's Non-Sensory Data} % title of Table
\centering % used for centering table
\begin{tabular}{|l|l|} % centered columns (4 columns)
\hline\hline %inserts double horizontal lines
Non-sensory data & Description \\
[0.5ex]
\hline
Timezone (TZ) &  Specifies the device's current timezone (i.e., a region
including the \\
& cities/states that have the same time) \\[0.5ex]
\hline
Device's address (IP) & Provides the phone's IP address when it is connected to the Internet\\ [0.5ex]
\hline
Network status (NS) & Specifies whether the smartphone is connected to a WiFi or a cellular network\\ [0.5ex]
\hline %inserts single line
\end{tabular}
\label{table:nonsensor}
\end{table*}

\begin{table*}[t]
\caption{Smartphone's Sensory Data} % title of Table
\centering % used for centering table
\begin{tabular}{|l |l |} % centered columns (4 columns)
\hline\hline %inserts double horizontal lines
Sensor & Sensory data\\
[0.5ex]
\hline
Accelerometer & Magnitudes of the acceleration in three-dimensional space\\[0.5ex]
\hline
Magnetometer  &  Angle between device's actual orientation relative to true north (heading) \\ [0.5ex]
\hline
Barometer  & The environment's air pressure\\ [0.5ex]
\hline %inserts single line
\end{tabular}
\label{table:sensor}
\end{table*}

\section{The proposed location mechanism}
\label{PMECH}
In this section, we describe PinMe, the proposed location mechanism.
First, we introduce the main sources of information that are given to
PinMe as inputs. Second, we describe various algorithms that we have
designed and implemented to locate the user in scenarios involving different activities. 

\subsection{Sources of information}
\label{SOURCE}
PinMe exploits two main sources of information: 
(i) non-sensory/sensory data collected by the smartphone, and 
(ii) publicly-available auxiliary information. Next, we describe each source 
in more detail. 

\subsubsection{Smartphone's non-sensory/sensory data}
An application installed on the smartphone can obtain several types of 
non-sensory and sensory data without requesting user's approval. Non-sensory 
data provide general information about the device, e.g., the version 
of the device's operating system, current timezone, IP address, the 
amount of available storage, and network status. Table \ref{table:nonsensor} 
summarizes different forms of non-sensory data that PinMe uses to locate the 
user during different activities, along with a short description of each. 

In addition to the non-sensory data, sensory data collected by the 
smartphone's built-in sensors provide valuable information about the user's 
movements and the environment in which the smartphone is located. 
Table \ref{table:sensor} includes different sensors that are accessed by 
PinMe and sensory data provided by each sensor.

\subsubsection{Publicly-available auxiliary information}
\label{INFOS}
The proposed user location mechanism uses several types of auxiliary 
information to narrow the area of interest. In particular, it utilizes five 
main types of information: (i) public maps, (ii) weather reports, (iii)
airports' specifications database, (iv) trains' heading dataset, and 
(v) transportation timetables. Next, we describe each information type.\\ \\
\noindent \textbf{Public maps:} The proposed mechanism uses two widely-known 
map types:

\noindent\textbf{1. Navigational map:} A navigational map mainly depicts roads, 
highways, and transportation links. Such a map can specify 
a large set of possible routes through which the user can travel. PinMe uses 
OpenStreetMap (OSM) \cite{OSM} maps. OSM maps can be downloaded as Extensible 
Markup Language (XML) files that can be easily processed and modified.\\
\noindent\textbf{2. Elevation map:} An elevation map contains the
elevation, i.e., the height above or below the Earth's sea level, of all points 
on the Earth's surface.  Several commercial, e.g., Google Map API \cite{GAPI}, 
and governmental services, e.g., U.S. Geological Survey Maps \cite{UAPI}, 
provide comprehensive 
elevation data of the world surface. For instance, the Google Map API offers 
a free and publicly-available interface that can be used by developers to 
fetch the elevation of a point of interest, given its longitude and latitude.\\ 

\noindent \textbf{Weather reports:} Weather reports offer different
types of information collected by weather stations. We use
weather reports provided by The Weather Channel \cite{WET}. They include 
temperature, humidity, and air pressure readings at weather stations, and 
the actual elevation of the weather station. PinMe uses weather reports to 
estimate the elevation of the smartphone using its air pressure reading. 
The use of weather reports is essential for accurately estimating the 
elevation of the smartphone since the air pressure readings are highly 
dependent on both elevation and weather conditions. \\

\noindent \textbf{Airports' specifications databases:} PinMe uses OpenFlights \cite{OFLIGHT}, the most comprehensive freely-available airports' specifications database, which includes elevation information, GPS coordinates, and timezone of 9541 different airports around the world.\\

\noindent \textbf{Trains' heading databases:} Trains' heading database is a 
simple database that includes the trains' directions at each station. 
We have constructed this database based on Google Map \cite{GOOGLEMAP}. For each train station considered in our experiments, we extract different potential movement directions based on the illustration of the stations' tracks on Google Map. Note that each track in a station can have two possible headings corresponding to a train entering and leaving the station.\\

\noindent \textbf{Transport timetables:} Transport timetables contain 
information about service times to assist passengers in planning their trip. 
A timetable lists the times when a service is scheduled to arrive 
(depart) at (from) specified locations\footnote{The actual 
destination/departure time may vary from the scheduled destination/departure 
time due to transportation delays. However, accurate information about the 
service is added to transport timetables after departure.}. The two most 
common types of transport timetables are flight and train timetables. 
These timetables are often available in a variety of 
electronic formats, e.g., PDF files, and are commonly posted on 
airports'/stations' websites. They are also accessible through various APIs. \\

\subsection{Main Algorithms}
Next, we describe the main algorithms that we have designed and implemented for estimating the user's location. PinMe is implemented using Python and Matlab, and our prototype implementation includes about 2000 lines of code. It has three main steps: (i) pre-processing, (ii) activity classification, and (iii) location estimation. \textit{Algorithm I: PinMe} provides a simplified pseudo-code of the proposed location mechanism. Next, we describe each step in more detail.\\

{
\noindent {\em Algorithm I: PinMe}\\
\noindent\makebox[\linewidth]{\rule{8.5cm}{0.1pt}}
\noindent Given: The smartphone's sensory data (D), non-sensory data (IP, NS, 
and TZ), and all sources of publicly-available auxiliary information 
(allAux: public maps, weather reports, airports' specifications 
databases, trains' heading databases, transport
timetables) \\
\noindent\makebox[\linewidth]{\rule{9cm}{0.5pt}}
\begin{center}
\vspace{-0.8cm}
\noindent\makebox[\linewidth]{\rule{9cm}{0.5pt}}
\end{center}
//Step 1: Pre-processing \\
\indent $lastWiFiIP \leftarrow findLastWiFiIP(NS,IP)$\\
\indent $city \leftarrow  IPGeolocation(lastWiFiIP)$\\
\indent $aux \leftarrow  getAux(allAux,city)$\\
\indent $chunks[] \leftarrow  streamPartitioning(D)$\\
//Step 2: Activity classification\\
\indent $acts[] \leftarrow  activityClassifier(chunks[])$\\
//Step 3: Location estimation \\
\indent \textit{for each activity in $acts[]$}\\
\indent \indent $[city, loc[i]] \leftarrow  Estimator(chunks[i],acts[i],aux,city)$\\
\indent \textit{end}\\
\indent  $return$ $loc[]$ \\
\noindent\makebox[\linewidth]{\rule{9cm}{0.5pt}}
\begin{center}
\vspace{-0.8cm}
\noindent\makebox[\linewidth]{\rule{9cm}{0.5pt}}
\end{center}

\subsubsection{Pre-processing}
In this step, PinMe first recognizes the last city in which the user was connected to a WiFi network and gets the required sources of auxiliary 
information for the potential city of interest. Second, it breaks the 
sensory data into several chunks so that each chunk is associated with a 
single activity.\\ 
\noindent \textbf{1. Inferring the city:} When the smartphone is connected to 
a WiFi network, IP geolocation techniques can process the device's current IP 
address and return the city in which the smartphone is located. Although such 
techniques can accurately locate the smartphone when it is connected to a 
WiFi network, they usually fail to locate it when connected to a cellular 
Internet network \cite{LF1,LF2}. 

Both iOS and Android allow an installed application to determine whether the 
smartphone is connected to a WiFi or a cellular network. In order to find the 
last city in which the user was connected to a WiFi network, PinMe processes 
the previous readings of smartphone's Network Status (NS) and IP address to 
find the last IP address of the smartphone when it was connected to a WiFi 
network, and feeds that IP address to \textit{$IPGeolocation(...)$}. Then, 
PinMe obtains different types of auxiliary information about the city, e.g., 
its maps.  \textit{PinMe does not assume that the user remains in the same 
city}. However, it starts tracking the user from that city. 
In fact, the user's current city becomes regularly updated 
based on his past movements.\\
\noindent \textbf{2. Data stream partitioning:} In the pre-processing
step, PinMe also breaks the long data stream collected over a long time
period, e.g., a day, into data chunks so that each chunk only includes the data associated with one activity. Based on our empirical analyses, a simple pattern in the acceleration data can indicate that a new activity 
has commenced: in the transition from one activity to another, the
accelerometer measures a series of large absolute acceleration readings 
(larger than 25 $m/s^2$) in a short time frame due to the fact that there is 
always a transition from standing (sitting) position to sitting (standing) 
position between two activities. This is the pattern PinMe uses to break the 
data stream into small data chunks. Unfortunately, a similar pattern might be 
present in the acceleration data collected during a single activity, e.g., 
when the user suddenly moves or falls. Therefore, it is possible that PinMe 
falsely detects the start of a new activity even when the user's activity has 
not changed. However, this does not negatively impact the accuracy of 
the location mechanism because as described later, for all activities, 
the activity classifier accurately detects the user's activity and PinMe can 
merge consecutive data chunks into one data chunk when the user's activity has 
not changed.   

\subsubsection{Activity classification}
\label{ACT_CLASS}
In this step, the activity classifier aims to specify the user's
activities. Throughout the paper, we assumed that the user takes part in one 
of the four activities mentioned earlier: driving, traveling on a plane, traveling on a train, and walking. To classify these activities, we have implemented two classification methods: (i) a traditional machine learning-based method that relies on building models to label the user's activities, and (ii) a tailored algorithm 
designed to deduce the user's activities based on the physical characteristics 
of each activity.

To the best of our knowledge, the activity classifiers utilized in PinMe are the first activity classification mechanisms that use air pressure data as a primary source of data for activity classification, and the first to use macro-level features, e.g., the number of turns and the rate of change during a turn, of heading data. Our examination of real-world data shows that air pressure and heading can offer valuable discriminatory information for activity classification. 

Fig.~\ref{fig:heading} illustrates how the smartphone's heading changes
in four data chunks collected during different activities. Among all
activities, traveling on a train is the only one in which the smartphone
observes \textit{no significant change in heading data}. Note that
heading data are measured clockwise from true north and vary from 
0$^{\circ}$ to 359$^{\circ}$. 

Fig.~\ref{fig:airpressure} shows how air pressure changes during different activities. Traveling on a plane is the only activity in which \textit{a fast significant drop in the environment's air pressure} was noticed. 

\begin{figure}[h]
\centering
\includegraphics[trim = 10mm 8mm 17mm 10mm ,clip, width=200pt,height=130pt]{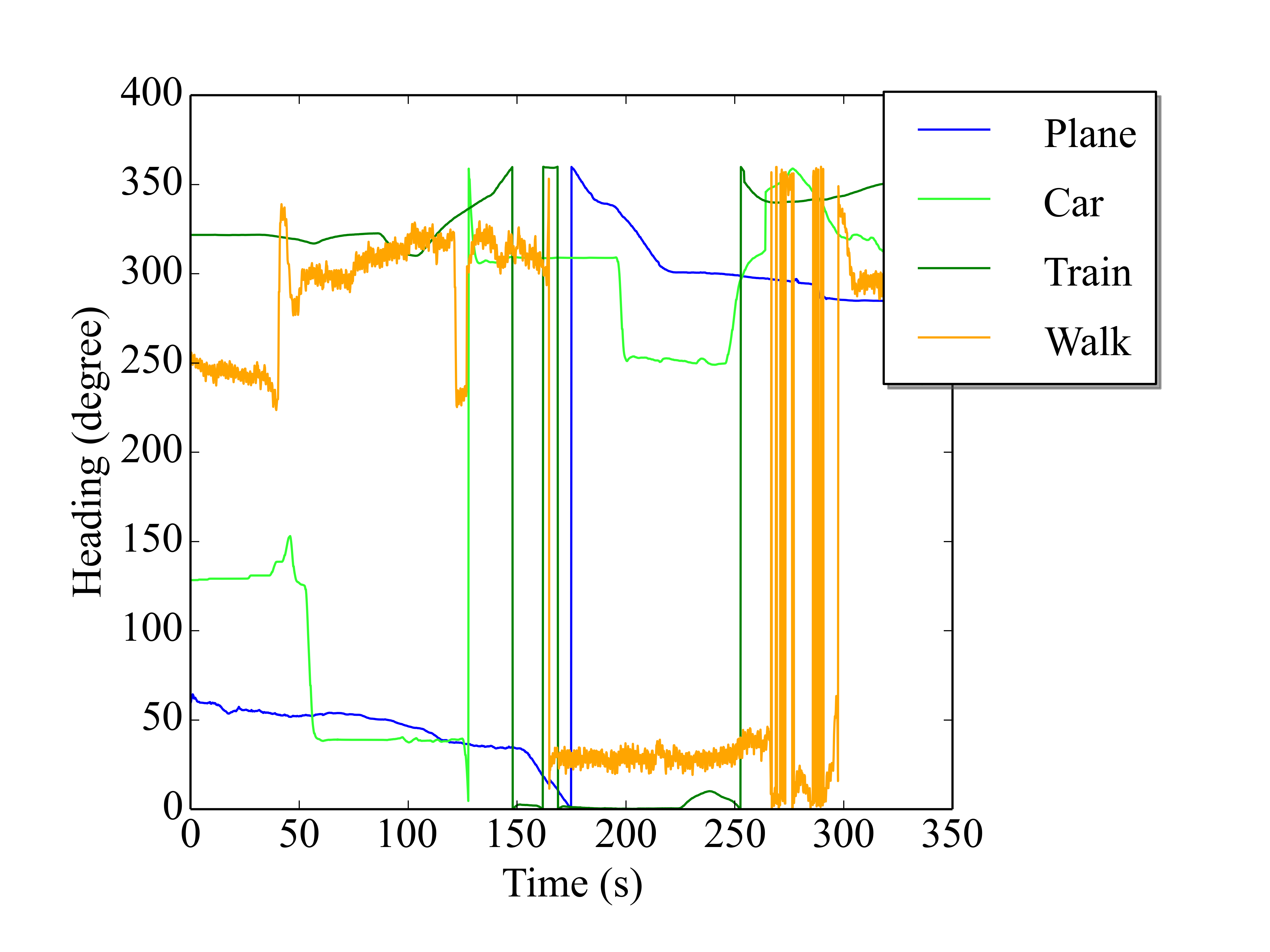}
\caption{Heading data collected during four different activities. Heading data are measured clockwise from true north and varies from 0$^{\circ}$ to 359$^{\circ}$. The smartphone's heading only slightly changes when the user is traveling on the train (within a 30-degree range).}
\label{fig:heading}
\end{figure} 

\begin{figure}[h]
\centering
\includegraphics[trim = 10mm 6mm 15mm 10mm ,clip, width=200pt,height=130pt]{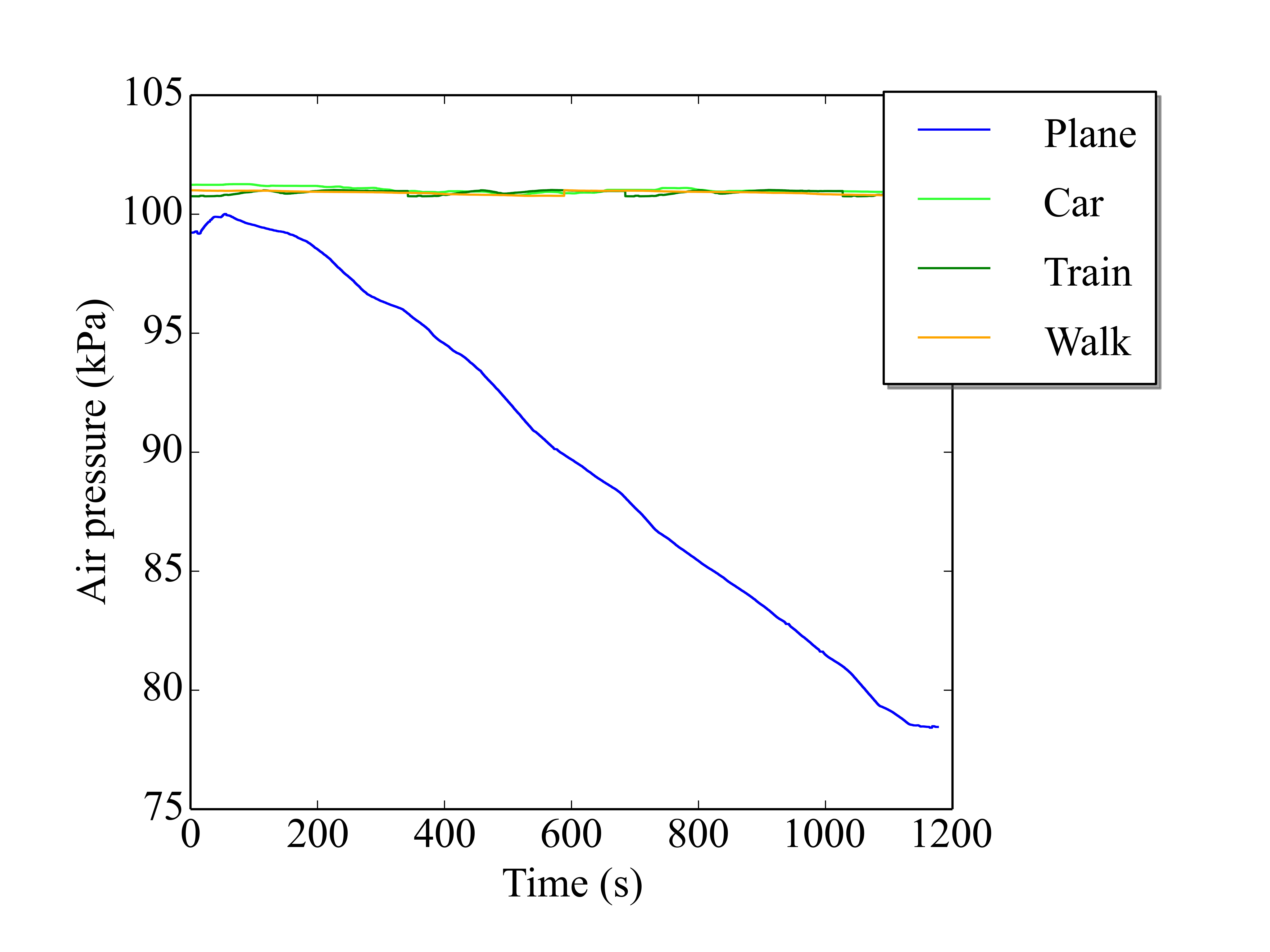}
\caption{Air pressure data collected during four different activities.}
\label{fig:airpressure}
\end{figure} 

Next, we describe each of the above-mentioned methods. \\
\noindent \textbf{Method 1: Machine learning-based classification}\\
\indent  A classical approach to implementing an activity classification 
mechanism is to devise a scheme based on a supervised machine learning 
algorithm, which builds a model using labeled training data. The
training dataset used for activity classification is not attack-specific (attacker can collect the required data using his own smartphone while traveling through unknown paths). This mechanism consists of three steps: feature extraction, 
binary classification, and decision making. Upon receiving a data chunk, the 
feature extraction step generates a feature vector. This vector is then sent 
to four binary classifiers, each trained to only detect a single activity. 
Finally, the decision making step returns the user's activity based on the 
outputs of the binary classifiers. Next, we discuss how each of these steps 
is implemented in our proposed scheme.\\
\noindent\textbf{1. Feature extraction:} Previous research efforts
\cite{ACTF1,ACTF2,ACTF3,ACTF4} have suggested a variety of features that
can be extracted from acceleration data and be used to classify various
user activities. In our mechanism, we use several features extracted
from heading and air pressure data along with a few previously-proposed
acceleration-related features. Each feature vector includes: time-domain features (mean, median, and standard deviation) and frequency-domain features (principal frequency and spectral energy) extracted from each dimension of acceleration readings, time-domain features (mean, median, and standard deviation, and range) from air pressure, and macro-level features (number of turns and maximum rate of change in heading over 1-second windows) from magnetometer readings.\\
\noindent\textbf{2. Binary classifiers:} In order to implement binary 
classifiers, we use Linear Support Vector Machine (LSVM) \cite{LSVM}. LSVM is 
one of the simplest, yet powerful, binary classification methods. The basic 
concept behind an LSVM is to find a hyperplane that separates the 
$n$-dimensional data 
into two classes. When no prior knowledge about the dataset is available, 
LSVMs usually demonstrate promising results and generalize well. They 
construct a decision boundary with the largest possible distance to data 
points. The binary classifiers used in the proposed scheme are trained so 
that each classifier can only recognize a single activity.\\ 
\noindent\textbf{3. Final decision making:} The final decision making step 
receives the 
classifiers' outputs, and returns an output as follows: if only one classifier 
detects the activity, it returns the activity associated with that classifier, 
otherwise, it returns a message stating that the activity is not recognized.

\noindent \textbf{Method 2: Tailored algorithm}\\
In addition to the machine-learning based method, we have developed a simple, 
yet accurate, classification algorithm. The simple tailored algorithm 
classifies the user's activities based on each activity's physical 
characteristics. We examine several data streams collected by the smartphone 
during different user activities. For each activity, we extract a set of 
characteristics that only pertains to that activity. 
Table \ref{table:PHYSICAL} summarizes these characteristics. 

\begin{table*}[t] 
\caption{Discriminatory characteristics of each activity} % title of Table 
\centering % used for centering table 
\begin{tabular}{|l|l|} % centered columns (4 columns) 
\hline\hline %inserts double horizontal lines 
Activity & Characteristics \\ [0.5ex]
\hline 
Driving & Irregular positive (negative) accelerations as the driver accelerates (brakes) \\ [0.5ex]
& Specific changes (around 90 degrees) in the smartphone's heading as the car turns\\ [0.5ex]
\hline 
Traveling on a plane & Rapid changes in the timezone \\ [0.5ex]
& Significant increase/decrease of air pressure in a short time frame\\ [0.5ex]

\hline 
Traveling on a train & Regular positive (negative) accelerations in one direction as the train leaves (reaches) a station \\ [0.5ex]
& No significant changes in the smartphone's heading\\ [0.5ex]

\hline 
Walking & Very frequent periodic acceleration changes in one direction, no matter how the device is held\\ [0.5ex]

\hline %inserts single line 
\end{tabular} 
\label{table:PHYSICAL}
\end{table*}

%\item Traveling on a plane: When the user is on a plane, his ground speed 
%(i.e., speed relative to the ground) is very high. This results in a possible 
%change in the device's timezone. In addition, the air pressure rapidly 
%increases/decreases as the aircraft decreases/increases its altitude. Although 
%the aircraft cabin is pressurized, the cabin's air pressure is lower than 
%the air pressure at sea level. For instance, the air pressure in the cabin is 
%equivalent to the outside air pressure at $1800$-$2400$ $m$ above sea level when 
%the aircraft is at a cruising altitude \cite{CAIR}.

%\item Traveling on a train: As a train leaves (reaches) a station, the 
%accelerometer observes a significant positive (negative) acceleration since 
%the train needs to immediately increase (decrease) its speed. Moreover, since 
%the train moves on a straight track, the changes in the train's 
%heading are negligible.

%\item Walking: When the user is walking, the accelerometer measures relatively large frequent periodic acceleration changes, no matter how the device is held. In fact, acceleration data contain a significant peak for each walking cycle (step). 

%\end{enumerate}

\subsubsection{Location estimation}
\label{ALGORITHMS}
In order to estimate the user's location, we have implemented four algorithms, 
referred to as location estimators. Upon detection of the user's activities 
($acts[]$) using the activity classifier, for each activity, PinMe calls 
$Estimator(...)$ that executes one of the four location estimators to find the 
user's locations. For each location estimator, Table \ref{table:inputOutput} 
summarizes the required non-sensory/sensory data and auxiliary information 
given to it and the outputs provided by each algorithm. Next, we describe the 
four proposed location estimators in more detail.  

\begin{table*}[t] 
\caption{The required non-sensory/sensory data and auxiliary information given to each location estimator and the outputs provided by each algorithm} % title of Table 
\centering % used for centering table 
\begin{tabular}{|l|l|l|} % centered columns (4 columns) 
\hline\hline %inserts double horizontal lines 
Location estimator & Inputs & Outputs\\ [0.5ex]
\hline 
Algorithm 1: carTracker & Air pressure, heading, public maps, & The initial and last locations and cities,\\ [0.5ex] 
& and weather reports & and the car's estimated trajectory\\ [0.5ex] 
\hline 
Algorithm 2: planeTracker & Air pressure, acceleration, TZ, weather reports,  & The destination and departure airports \\ [0.5ex]
& airports' specifications databases, & \\ [0.5ex]
& and flight timetables & \\ [0.5ex]
\hline 
Algorithm 3: trainTracker & Acceleration, heading, train timetables, & The destination and departure stations\\ [0.5ex]
& and trains' heading databases &\\ [0.5ex]
\hline 
Algorithm 4: walkingUserTracker & Air pressure, acceleration, heading, weather reports, and public maps & The user's last location and trajectory\\ [0.5ex]

\hline %inserts single line 
\end{tabular} 
\label{table:inputOutput}
\end{table*}

\noindent\textbf{Algorithm 1: carTracker:} Unlike the method in \cite{ACCO} that uses very noisy accelerometer measurements, this algorithm relies on the sensory data collected by the magnetometer and barometer (heading and air pressure) to provide a very accurate tracking mechanism. It has three main steps:

\textit{Step 1: Map construction:} Prior to tracking the user, PinMe constructs a labeled directed graph $G$ using both elevation and navigational maps of the 
city so that its vertices and edges represent the intersections and roads between intersections, respectively. Labels of vertices are the elevation of the intersections extracted from the navigational map and the angle between roads connecting to that intersection. 

\textit{Step 2: Pruning set of probable candidates:} At each moment of
time, the algorithm has an array of trees (the set of probable paths
with different starting points, referred to as $P$) where each tree represents 
a sequence of intersections on the navigational map. Prior to the attack, this 
array contains all vertices of $G$, indicating that the first turn can be at 
any intersection. Upon the detection of a turn (e.g., an almost 90-degree 
change in the heading data), the algorithm prunes and updates the set of trees 
as follows. For each probable path (each tree in set $P$), it drops the path 
if all neighbors of its last vertex do not meet the following conditions: the 
elevation or relative changes in the heading direction of all neighbors 
(represented as labels of vertices in graph $G$) do not match their values 
extracted from sensory data.

\textit{Step 3: Updating the remaining candidates:} At each turn, if a tree is not dropped from the set, the algorithm adds all neighbors (intersections) that meet the above-mentioned conditions to the tree. Eventually, it sorts paths in $P$ based on their error, defined as the weighted sum of absolute differences between the extracted features from the sensory data and their actual values reported in navigational/elevation data, and returns the most probable path from the set (the path with the lowest error). 

Although the number of intersections of a city is large, we observe based on 
experimental results that \textit{the number of intersections that can be a 
part of a candidate path drops extremely fast from thousands to only a 
few after the first few turns}. As a result, the size of set $P$ is reduced 
quickly as the algorithm removes many impossible candidates when they become 
inconsistent with new data. This is demonstrated later in Section \ref{EVAL}.

\noindent \textbf{Note:} Although there is a well-known physics equation 
\cite{S_FORMULA} for estimating elevation (relative to sea level) based on 
air pressure measurements alone, it does not provide an accurate estimation of 
the elevation in practice since barometer measurements significantly depend 
on weather conditions. To accurately estimate the elevation ($H_{turn}$) of 
a turn point, given the air pressure measured at the point ($P_{turn}$), 
PinMe first extracts the air pressure ($P_{station}$), elevation 
($H_{station}$), temperature information ($T$), and humidity (indicated by a 
constant $C$) at $city$'s weather station, provided by its weather
report, and then uses the following physics equation \cite{FORMULA}:
\begin{equation}
    H_{turn}=H_{station} + \frac{T}{C} ln(\frac{P_{turn}}{P_{station}})
\end{equation}

%\textit{To sum up, the carTracker algorithm uses air pressure and heading readings from the smartphone, weather reports, and both elevation and navigational maps of the city.}

\noindent\textbf{Algorithm 2: planeTracker:} \textit{planeTracker} first 
extracts three features from the raw data provided by the smartphone: 
(i) flight time data (takeoff and landing times and flight duration), 
(ii) TZ and elevation of the departure airport, and (iii) TZ and elevation 
of the destination airport. In order to extract these features from the raw data, 
the algorithm first recognizes different aviation phases of the flight 
(pre-flight, takeoff, cruising, descending, landing, and taxiing to the gate) 
by processing acceleration and elevation data collected by the smartphone 
during the flight. Then, it calculates the flight duration as the time 
difference between the pre-flight phase (i.e., when the plane leaves the gate 
at the departure airport) and taxiing phase (i.e., when the plane reaches 
the gate at the destination airport). Moreover, it stores the device's air 
pressure and TZ in both the pre-flight and taxiing phases. Afterwards, it 
calculates the elevations of both departure and destination 
airports, given the weather report (including the air pressure reading at 
$city$'s weather station and its elevation data). Then, it 
searches through the airports' specifications database to find the
flight routes, which have the following characteristics: (i) the TZ of both destination and departure airports reported by the smartphone matches the ones reported in the database, (ii) the difference between elevation measured from air pressure data and elevation extracted from the database is less than a small threshold, e.g., $T_{elevation}=5m$, and (iii) the difference between flight duration measured from acceleration data and flight duration extracted from the database is less than a certain threshold, e.g., $T_{duration}=1h$. 

Given timetables of probable departure/destination airports,
\textit{planeTracker} returns the routes for which both takeoff time and
landing time almost match their corresponding times provided by
timetables, e.g., $\Delta T_{landing}, \Delta T_{takeoff} <1h$, where
$\Delta T_{landing/takeoff}$ is the difference between landing/takeoff
times extracted from sensory data and their expected values in timetables.

\noindent\textbf{Algorithm 3: trainTracker:} Acceleration data can reveal different transportation phases, e.g., when the train leaves or approaches a station, and the combination of acceleration and heading data provides an approximation of the train's heading. This algorithm has two main steps:

\textit{Step 1: Extracting features:} It first extracts three features from the raw acceleration and heading data: (i) travel intervals (an array $T$), 
defined as the difference between the time the train leaves a station and the time it reaches the next station, (ii) departure time $T_{departure}$ that represents when the train left the first station, and (iii) train's heading, i.e., an approximation of the direction of the train's movement at the first station.

\textit{Step 2: Searching through the timetable:} After extracting the above-mentioned features from the raw data, this algorithm searches the timetables of $city's$ stations to find the most probable route. It first constructs 
$T_{train}$ for all trains that already left or will leave the current city around the departure time (within $T_{departure}-1h$ to $T_{departure}+1h$) as follows: each $T_{train}$ is itself an array including travel intervals for a single train. Then, for each $T_{train}$ in the list, it computes the difference between travel intervals extracted from the sensory data ($T$) and $T_{train}$, i.e., $D=\sum_{n=1}^{length(T)} |T[i]-T_{train}[i]|$. If the difference between $T$ and $T_{train}$ is below a certain threshold (i.e., $D < 2mins \times length(T)$), the route corresponding to $T_{train}$ is added to the set of probable routes ($P$). Then, the algorithm prunes $P$ by removing routes for which the difference between the trains' heading extracted from the sensory data and the actual value of heading reported in trains' heading database is above a certain threshold (30 degrees). Finally, from the remaining routes, it returns a single route corresponding to the lowest $D$ in the set.

\noindent\textbf{Algorithm 4: walkingUserTracker:} 
This algorithm assumes that the user walks through the walking areas (roads or sidewalks) of the navigational map. We have implemented two different versions of the algorithm. The first version searches through the whole map to find the user's trajectory. However, to find the initial location of this activity, the second version only considers a small area ($300m \times 300m$) on the map around a given location (in real-world scenarios, this location is determined by a previous activity). Next, we describe the first version that has three steps (the second version is similar, however, it only considers a smaller set of nodes to find the initial point).

\textit{Step 1: Map construction:} Prior to the attack,
\textit{walkingUserTracker} constructs a graph $G$ similar to the one
generated for \textit{Algorithm 1: carTracker}, with a slight
difference: the graph also has a label on each edge that represents the
length of the corresponding road extracted from the navigational map.
Similar to \textit{carTracker}, the algorithm maintains an array of
trees (the set of probable paths with different starting points, referred to 
as $P$) where each tree represents a sequence of intersections on the 
navigational map. 

\textit{Step 2: Pruning the set of probable candidates:} The algorithm extracts the steps and their direction from the raw acceleration and heading data and elevation of intersections from air pressure readings. Upon the detection of a turn (e.g., an almost 90-degree change in the heading data), the algorithm updates the set of trees as follows. For each probable path, it drops the path if all 
neighbors of its last vertex do not meet at least one of the following conditions: (i) all labels of edges that connect the last vertex to it neighbors ($D[i]$s) do not match the estimation of the travelled distance calculated based on the number of steps (for example, all $D[i]$s are not within the range of $0.4m\times\#steps$ to $1.2m\times\#steps$), or (ii) the elevation or relative changes in heading direction of neighbors do not match their values extracted from sensory data.

\textit{Step 3: Updating the remaining candidates:} At each turn, if a
tree is not eliminated, the algorithm extends it by adding all neighbors (intersections) that meet the above conditions. This algorithm sorts paths $P$ based 
on their error, defined as the weighted sum of absolute differences between the extracted features from the sensory data and their actual values given by maps, and returns the path with the lowest error. 

\textbf{Note:} Although this algorithm uses an estimation of the distance walked by the user to find the trajectory, it can also accurately estimate the user's step size upon the detection of a unique path. It uses the information gathered in the last sidewalk/road (e.g., total number of steps) along with information offered by the navigational map (e.g., the total length of the last sidewalk/road) to adaptively estimate the user's step size. Upon the detection of a unique trajectory, the estimation of the step size enables the algorithm to accurately estimate the user's location on the road. 

\section {Evaluation of the proposed mechanism}
\label{EVAL}
In this section, we first describe our data collection procedure. Then, we examine the 
accuracy of PinMe using real-world data.

\subsection{Data collection procedure}

We start with the description of the data collection procedure.
\subsubsection{Device characteristics and experimental configurations}
The proposed location mechanism is evaluated on three smartphones (Galaxy S4 
i9500, iPhone 6, and iPhone 6S). Each device is equipped with an internal GPS 
device and several high-precision sensors including, but not limited to, 
a 3/6-axis accelerometer, magnetometer, and barometer.

As mentioned earlier in Section \ref{PMECH}, PinMe processes various
types of sensory data (air pressure, heading, and acceleration) and
non-sensory data (the device's TZ, IP, and NS). In order to collect the 
required data using Galaxy S4 i9500, we developed an Android application that 
continuously records the non-sensory/sensory readings of the device. Moreover, 
we installed a sensor data logger application on both iPhone 6 and iPhone 6s, 
called SensorLog \cite{SENSORLOG}, which continuously records the required 
non-sensory/sensory data. In our data collection procedure, sensory data are collected at the sampling frequency of $5 Hz$. In addition to the above-mentioned data, the applications installed on the smartphones also collect GPS readings. GPS data are only used to evaluate the 
accuracy of PinMe in estimating the user's location (PinMe does not access GPS 
data).

\subsubsection{Datasets}
\label{DATASET}
We constructed two datasets using real-world data. The first dataset consists 
of several data chunks, i.e., sequences of consecutive readings of 
non-sensory/sensory data collected during one activity. The second dataset 
includes three non-sensory/sensory data streams collected by the three 
under-experiment smartphones for a whole day. Next, we briefly describe each 
dataset. During the collection of each data chunk, the smartphone's 
orientation was almost fixed, however, the actual orientation of the smartphone was 
unknown in all cases.\\\\
\noindent\textbf{Dataset \#1}: This dataset consists of 405 data chunks 
collected during different user activities where each data chunk contains 
consecutive readings of air pressure, heading, acceleration, and the device's 
TZ, IP, and NS during each activity. Table \ref{table:sum} shows the number of 
collected chunks for each activity. Next, for each activity, we briefly 
describe how we collected real-world data.

\begin{table}[ht] 
\caption{Number of data chunks in Dataset \#1 for each activity} % title of Table 
\centering % used for centering table 
\begin{tabular}{l c} % centered columns (4 columns) 
\hline\hline %inserts double horizontal lines 
Activity & Number of data chunks \\ [0.5ex]
\hline 
Driving & 271\\ [0.5ex]
Traveling on a plane & 4\\ [0.5ex]
Traveling on a train & 30\\ [0.5ex]
Walking & 100\\ [0.5ex]
\hline %inserts single line 
\end{tabular} 
\label{table:sum}
\end{table}

\noindent\textbf{1. Driving:} A user, carrying an iPhone 6, drove in three different cities. 271 data chunks were collected, where each chunk contains the smartphone's data during one driving period. Table \ref{table:DN} shows the cities in which the user drove, their populations, the state in which each city is located, and the number of collected data chunks for each city. To provide a fair evaluation, we tried to collect data chunks from different areas of these cities (both dense and sparse areas).

\begin{table}[ht] 
\caption{Cities, their populations and state, and the number of driving chunks 
for each city} % title of Table 
\centering % used for centering table 
\begin{tabular}{l c c c} % centered columns (4 columns) 
\hline\hline %inserts double horizontal lines 
City name & Population & State & Chunks \\ [0.5ex]
\hline 
Princeton & 12307 & NJ & 105\\ [0.5ex]
Trenton & 84308 & NJ & 111\\ [0.5ex]
Philadelphia & 1.5 million & PA & 55 \\ [0.5ex]
\hline %inserts single line 
\end{tabular} 
\label{table:DN}
\end{table}

\noindent \textbf{2. Traveling on a plane:} We collected four data chunks when 
the user traveled on four different airplanes on four different flight 
routes: (i) from Philadelphia to Dallas, (ii) from Dallas to New York, (iii) 
from College Station to Dallas, and (iv) from Dallas to College Station. All 
four data chunks were collected using iPhone 6S.  \\
\noindent\textbf{3. Traveling on a train:}  We collected 30 data chunks using an 
iPhone 6s when the user traveled on a train (10 chunks for Princeton Junction 
Station to New York, 10 chunks for Baltimore Penn Station to New York, and 10 
chunks for Washington D.C. Union Station to New York).\\
\noindent\textbf{4. Walking:} We collected 100 data chunks when the user walked 
carrying an iPhone 6. These data chunks were gathered in Princeton.  \\\\
\noindent \textbf{Dataset \#2:} This dataset includes three data streams 
collected by three users while going through their regular daily activities. 
Two users were located in Princeton, NJ and one user was located in Baltimore, 
MD. In order to construct this dataset, we asked the users to choose and carry 
one of the three under-experiment smartphones (Galaxy S4 i9500, iPhone 6, and 
iPhone 6S).       

\subsection{Accuracy evaluation}
In the following, we first evaluate the accuracy of the two main steps of PinMe (activity classification and location estimation) using Dataset \#1. Then, we use Dataset \#2 to provide an 
end-to-end evaluation.

\subsubsection{Step-by-step evaluation}
Next, we evaluate the accuracy of the activity classifier and location 
estimators using Dataset \#1.

\noindent \textbf{Evaluating the activity classifier}\\
We evaluated the two activity classification methods discussed in 
Section \ref{PMECH} using Dataset \#1. In the machine-learning based approach, we used $50\%$ of the collected data chunks for 
training the binary classifiers, and tested the accuracy of the scheme using 
data not used in the training phase. In the other approach, we 
used all data chunks to test the accuracy of the tailored algorithm. 
Both methods provided a classification accuracy of $100\%$, where 
classification accuracy is defined as the ratio of correctly recognized 
activities to the total number of activities processed by the activity 
classifier. A high classification accuracy was expected since each of the supported activities (driving, traveling on a plane, traveling on train, and walking) has unique physical characteristics that differentiate it from other activities.

\noindent \textbf{Evaluating the location estimators} \\
Next, we examine how accurately the four location estimator algorithms discussed in Section \ref{PMECH} can estimate the user's location.

\noindent\textbf{Algorithm 1: carTracker:}
In order to evaluate the accuracy of \textit{carTracker}, we used 271 data 
chunks from Dataset \#1, which were collected in three different cities 
(Table \ref{table:DN}). Next, we examine how 
accurately this algorithm can locate the user when it returns the most probable 
driving path from the set of probable driving paths and how the size of the 
set changes with respect to the length of the driving path.

Fig.~\ref{fig:DR1} shows the average approximation error with respect to
the length of the driving path, i.e., the number of routes the driver 
traverses in one driving period that is equal to the number of turns plus one. 
The approximation error is defined as the distance between the actual location 
(as provided by GPS sensor) and the estimated location (as estimated by PinMe) 
of the user, divided by the total traveled distance (computed by processing GPS 
readings). In our experiments, the length of the driving path varies between 5 
and 18. As can be seen from this figure, as the length of the driving path 
increases, the approximation error of the estimator typically decreases. 

\begin{figure}[h]
\centering
\includegraphics[trim = 8mm 0mm 20mm 10mm ,clip, width=200pt,height=130pt]{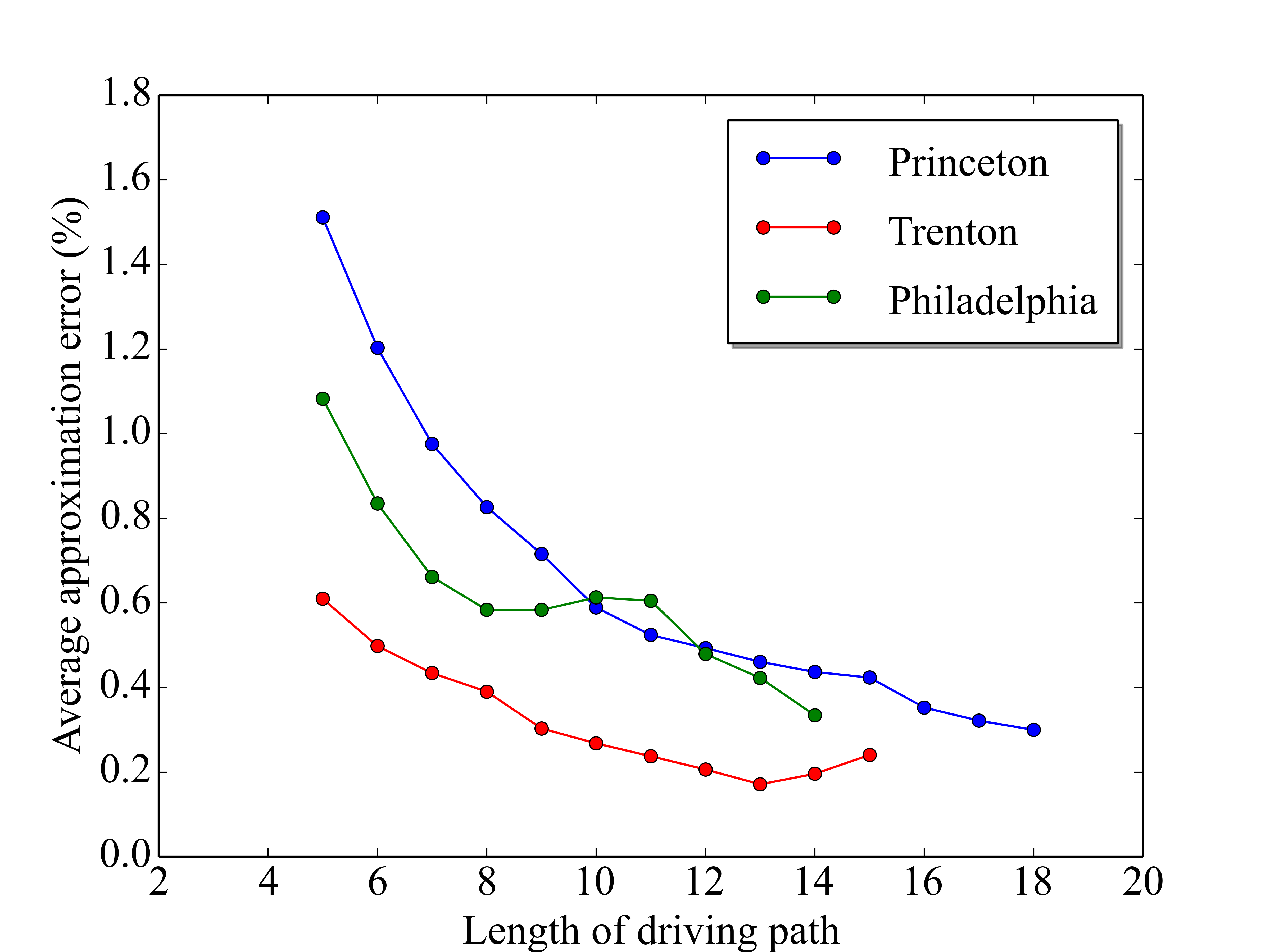}
\caption{Average approximation error with respect to the length of the driving 
path. The average approximation error is less than $1.5\%$ in all cases.}
\label{fig:DR1}
\end{figure} 

We examined how the number of possible driving paths decreases when the length 
of the driving path increases. Fig.~\ref{fig:DR2} illustrates the number of 
possible driving paths with respect to the length of the driving path. 
As can be seen, the number of possible driving paths drops 
rapidly as the length of the driving path increases. 

To sum up, as the length of the driving path increases, PinMe collects more 
information about the user's environment, and as a result, it is more likely 
to find a unique driving path on the map. 

\begin{figure}[h]
\centering
\includegraphics[trim = 6mm 0mm 20mm 10mm ,clip, width=200pt,height=130pt]{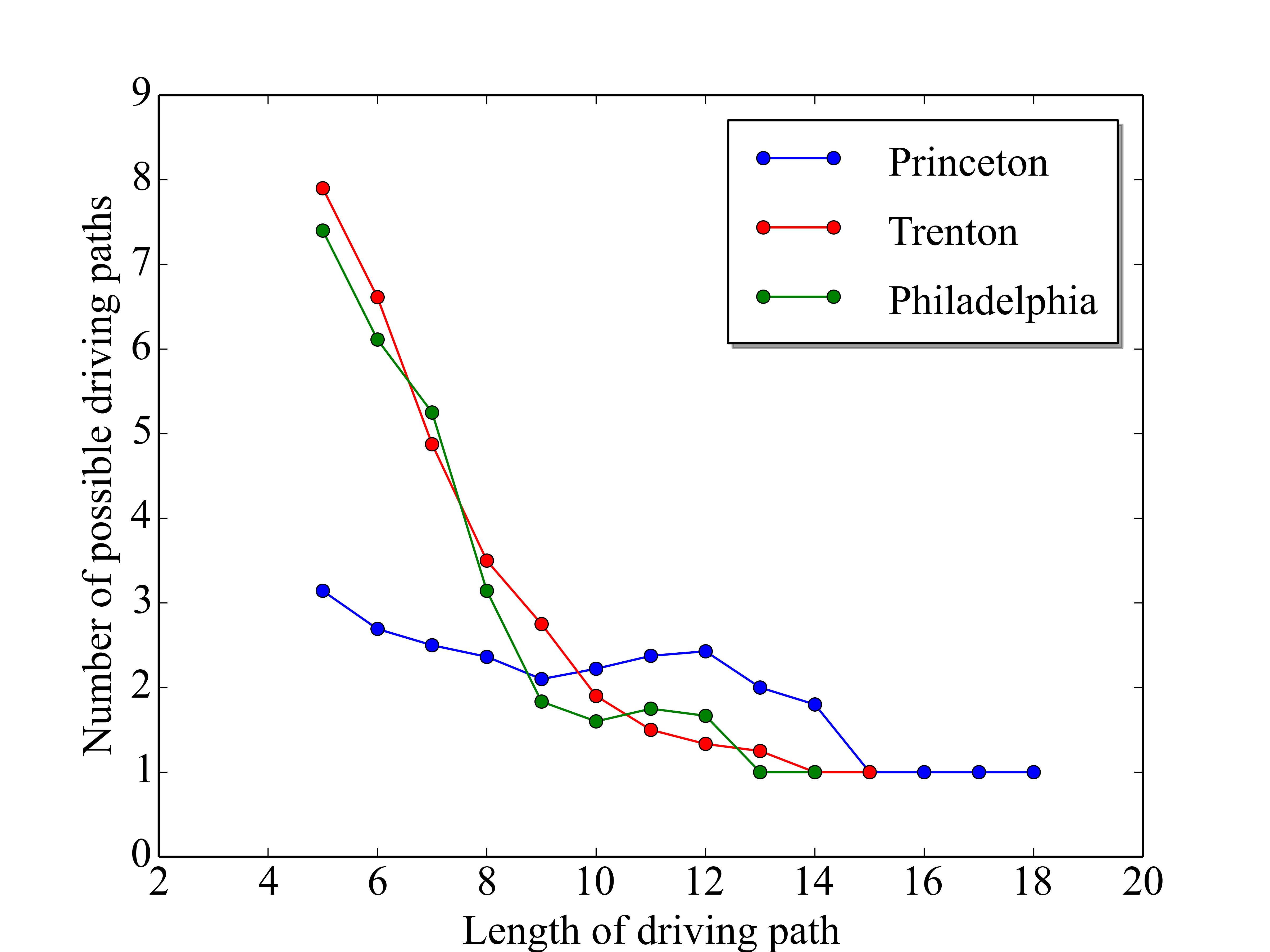}
\caption{Number of possible driving paths with respect to the length of
the driving path}
\label{fig:DR2}
\end{figure} 

%MOVE UP
%Since our tracking algorithm for driving analysis is based on route pattern match, there might be one or more routes satisfying our searching requirements.  In our experiment, we only add the turns which satisfy the following conditions as the next possible turns:
%\begin{enumerate}
%    \item angle error smaller than $\theta$ ,
%    \item altitude error smaller than $h$m,
%   \item vehicle speed must be reasonable.
%\end{enumerate}
%We pick up $\theta = 20^{\circ}$, $h = 2.0$, and $speed < 60mph$ in our experiments, and the number of routes returned from our application is shown in Figure \ref{fig:DR2}.

\noindent\textbf{Algorithm 2: planeTracker:} We examined the accuracy of 
\textit{planeTracker} in finding departure and destination airports using 
Dataset \#1. As shown in Table \ref{table:sum}, we collected four data chunks 
while traveling on a plane. Despite the existence of potential differences 
between the approximated values of takeoff time, landing time, and elevation, 
and their expected values reported in airports' specification database and 
flight timetables, \textit{planeTracker} was able to \textit{accurately and uniquely} return both departure and destination airports for all four flight routes. 

For each of the four data chunks, we examined how much the approximated
takeoff time, landing time, and elevation readings extracted by
processing the smartphone's sensory data differ from their expected
values calculated by processing publicly-available auxiliary data
(airports' specification database and flight timetables), and noticed
that: (i) the average difference between estimated elevation 
reported by the smartphone and the elevation extracted from
airports' specification database was $2.3$ $m$, (ii) the average
difference between the estimated flight duration and the actual flight
duration was $4\%$ of the actual duration, (iii) the difference
between approximated takeoff time and the takeoff time reported in the
flight timetable (flight delay) was 17 minutes. 

In addition to the above-mentioned analyses, we 
also examined the discriminatory power of the features extracted by 
\textit{planeTracker} (flight duration, TZs, and elevations of both destination and 
departure airports) using Monte Carlo simulation methodology
\cite{MONTE}. We considered two scenarios: (i) similar to
above-mentioned real-world cases, both departure and destination
airports are unknown and \textit{planeTracker} returns the flight route
(departure and destination airports), and (ii) attacker knows the
departure airport from a previous activity, e.g., driving to the
airport, and he only wants to identify the destination airport. For each scenario, we generated 500 random flight routes assuming that (i) for each route, the difference between the estimated flight duration and actual flight duration 
varies between $0\%$ and $10\%$ of the actual duration, and (ii) the 
difference between the estimated elevation reported by the smartphone 
and the elevation extracted from airports' specification database 
varies between $0$ $m$ and $5$ $m$. We slightly modified $planeTracker$ so that it returns the three most probable flight routes using the extracted features (without even using flight timetables). After finding a set of probable flight routes, it sorts the routes based on their error, defined as the weighted sum of absolute differences between the features (elevation and flight duration) calculated from sensory data and their expected values extracted from airports' specifications database. 

Fig.~\ref{fig:both} demonstrates how accurately $planeTracker$ is able
to find the actual flight route without knowing the departure airport,
where accuracy is defined as the number of cases in which the actual
flight route was among the three returned flight candidates divided by
the total number of trials (500). Similarly, Fig.~\ref{fig:destination}
shows how accurately $planeTracker$ can find the destination airport, given the departure airport. Despite the presence of potential differences between the approximated duration and elevation and their expected values, in the majority of cases, \textit{planeTracker} was able to find a set of three routes/destination airports that includes the actual flight route/destination airport, as illustrated in Fig.~\ref{fig:both} and Fig.~\ref{fig:destination}, respectively.

\begin{figure}[h]
\centering
\includegraphics[trim = 0mm 0mm 0mm 15mm ,clip, width=200pt, height=130pt]{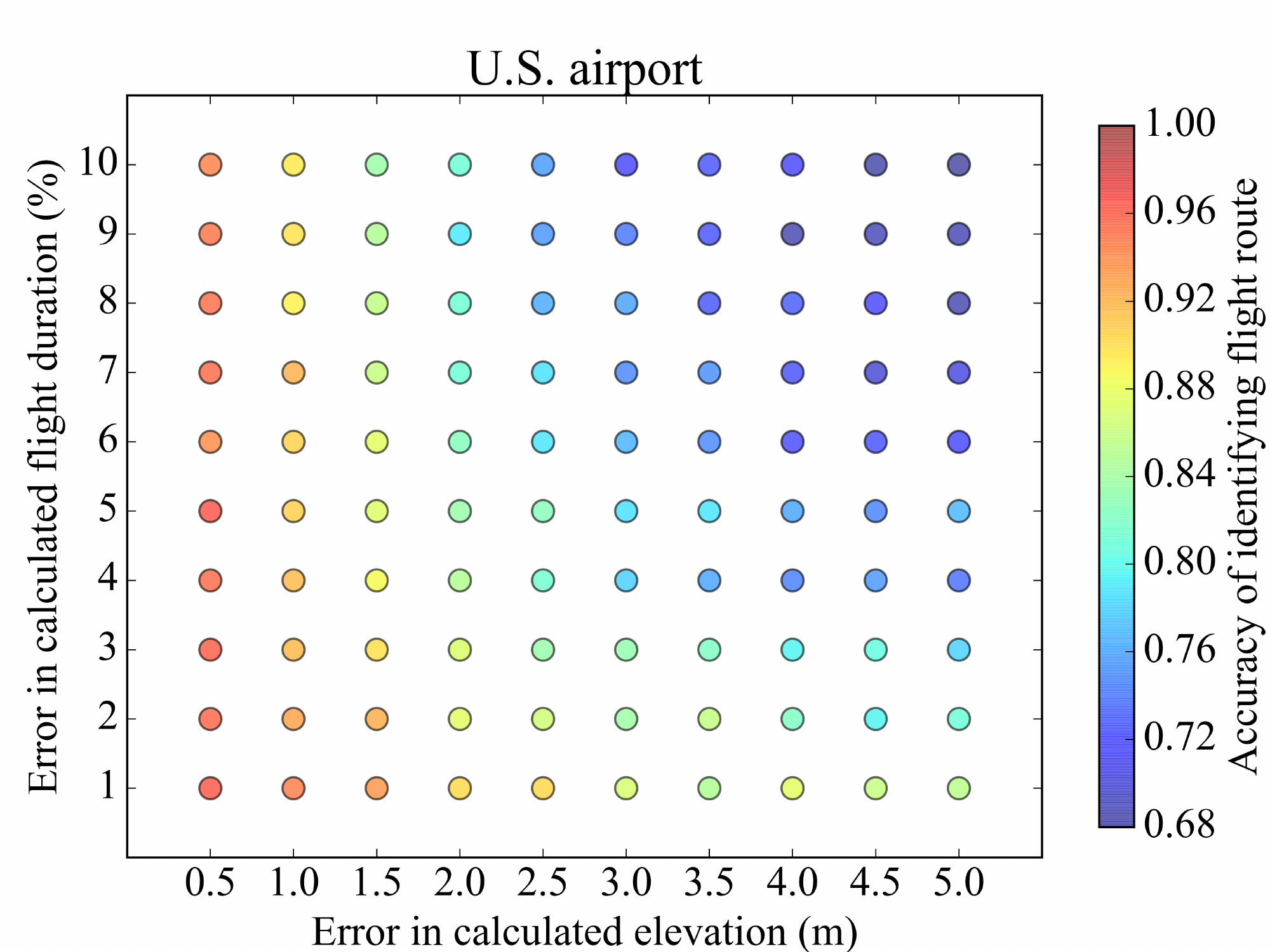}
\caption{Accuracy of $planeTracker$ in providing a set of three potential candidates so that the actual flight route is in the set.}
\label{fig:both}
\end{figure} 

\begin{figure}[h]
\centering
\includegraphics[trim = 0mm 0mm 0mm 15mm ,clip, width=200pt,height=130pt]{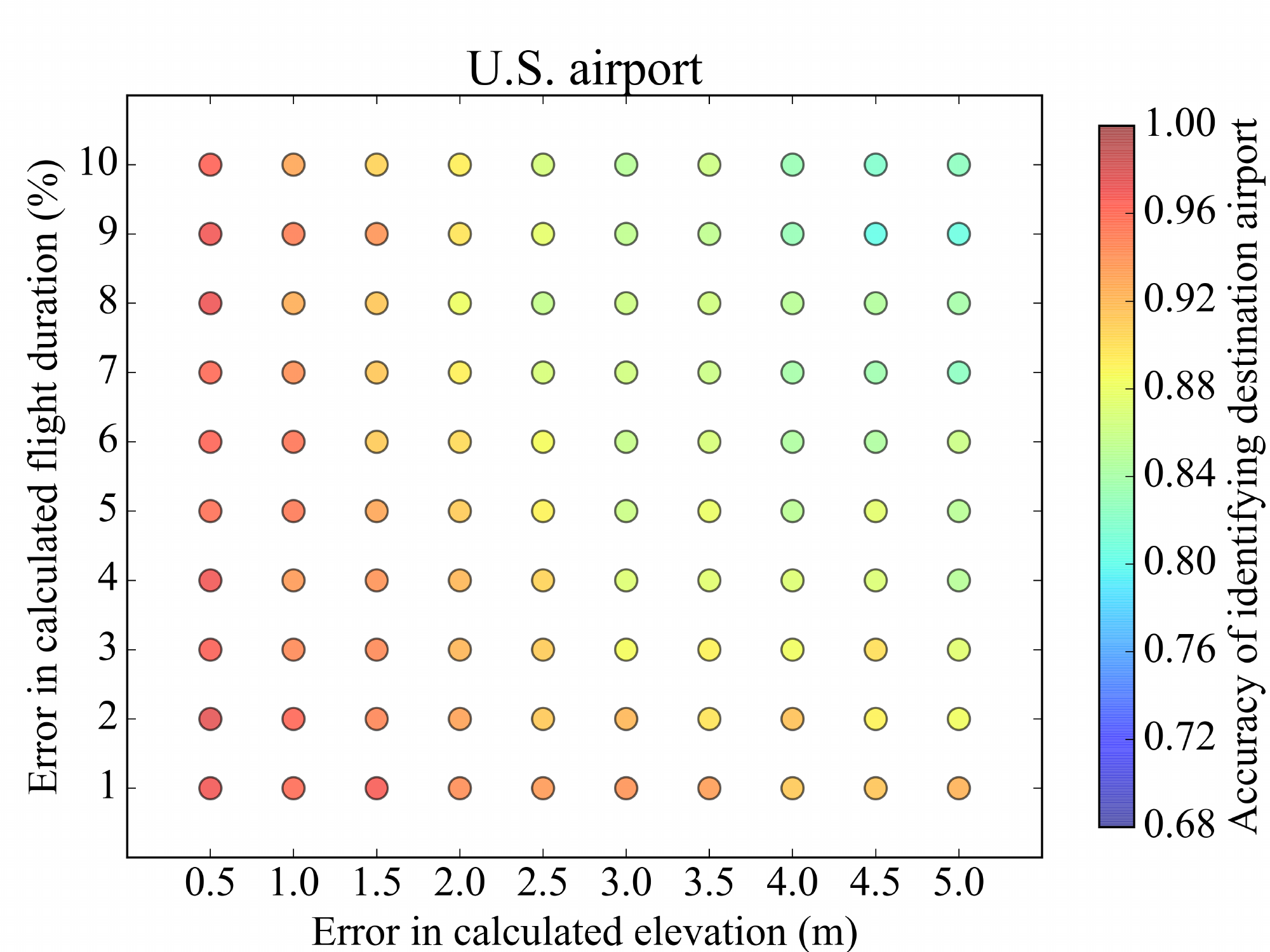}
\caption{Accuracy of $planeTracker$ in providing a set of three potential destination airports (given the departure airport) so that the actual destination airport is in the set.}
\label{fig:destination}
\end{figure} 

\noindent\textbf{Algorithm 3: trainTracker:}
As mentioned earlier, \textit{trainTracker} returns both departure and destination stations. We examined the accuracy of the tracking mechanism in 
finding actual traveling routes using the 30 data chunks collected by
the smartphone (10 chunks for Princeton Junction Station to New York, 10
chunks for Baltimore Penn Station to New York, and 10 chunks for Washington 
D.C. Union Station to New York). Our experimental results demonstrated that 
\textit{trainTracker} was able to accurately identify the user's travel route 
in all trials. 

\noindent\textbf{Algorithm 4: walkingUserTracker:}
As mentioned earlier, two different versions of \textit{Algorithm 4: walkingUserTracker} have been implemented: one that searches the whole map, and the other one that assumes the initial location is within a small area ($300m \times 300m$) around the final location of the last activity. Fig. \ref{fig:LONG} shows how the number of possible walking paths will change with respect to the number of walking steps for the first version of the algorithm. Based on our empirical results, although the possible number of candidates is reduced quickly, the possibility of each of them at each moment of time is similar to the others (i.e., when the number of steps is small, uniquely distinguishing the actual path is not feasible). As shown in Fig. \ref{fig:LONG}, in order to return a unique accurate path, the first version of the algorithm requires a long stream of sensory data (i.e., the user should walk over 2500 steps). We observed that, in real-world scenarios, users usually walk shorter distances (including only a few different roads) preceded by other activities (commonly driving). Thus, to accurately track the user in real-world scenarios during multiple activities, we suggest using the second version of the algorithm that utilizes the data provided by the previous activity.

\begin{figure}[h]
\centering
\includegraphics[trim = 0mm 0mm 10mm 10mm ,clip, width=200pt,height=130pt]{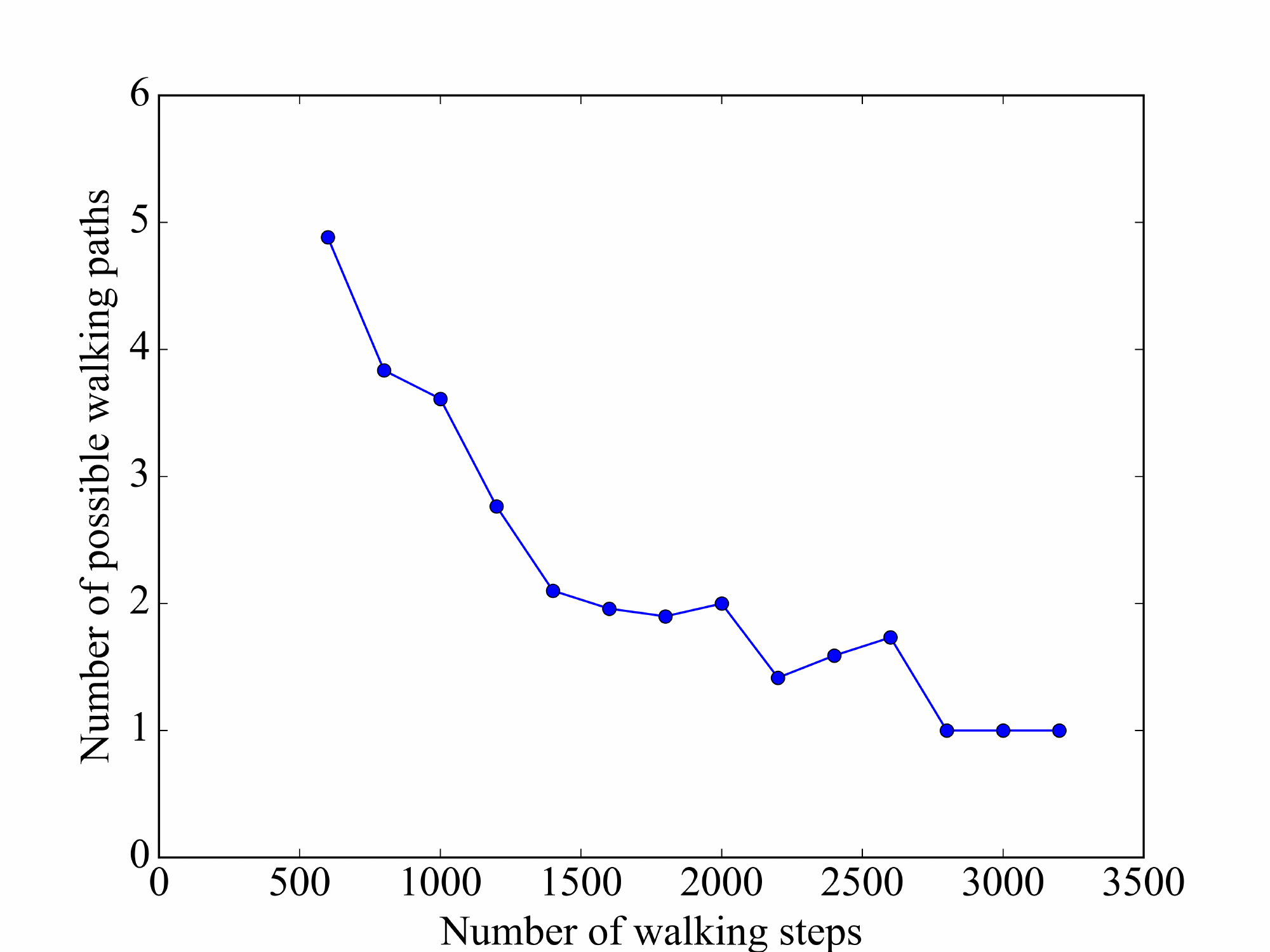}
\caption{Number of potential candidates with respect to the number of walking steps}
\label{fig:LONG}
\end{figure} 

We examined how accurately the second version of \textit{walkingUserTracker} estimates the user's location. Fig.~\ref{fig:walking} shows the approximation error for all walking trials with respect to the number of steps, where approximation error is defined as the distance between the 
user's actual location and the user's estimated location (as estimated by PinMe), divided by the total walking distance. 
As shown in the figure, the approximation error was less than $2.5\%$ for all data chunks.

\begin{figure}[h]
\centering
\includegraphics[trim = 0mm 0mm 10mm 10mm ,clip, width=200pt,height=130pt]{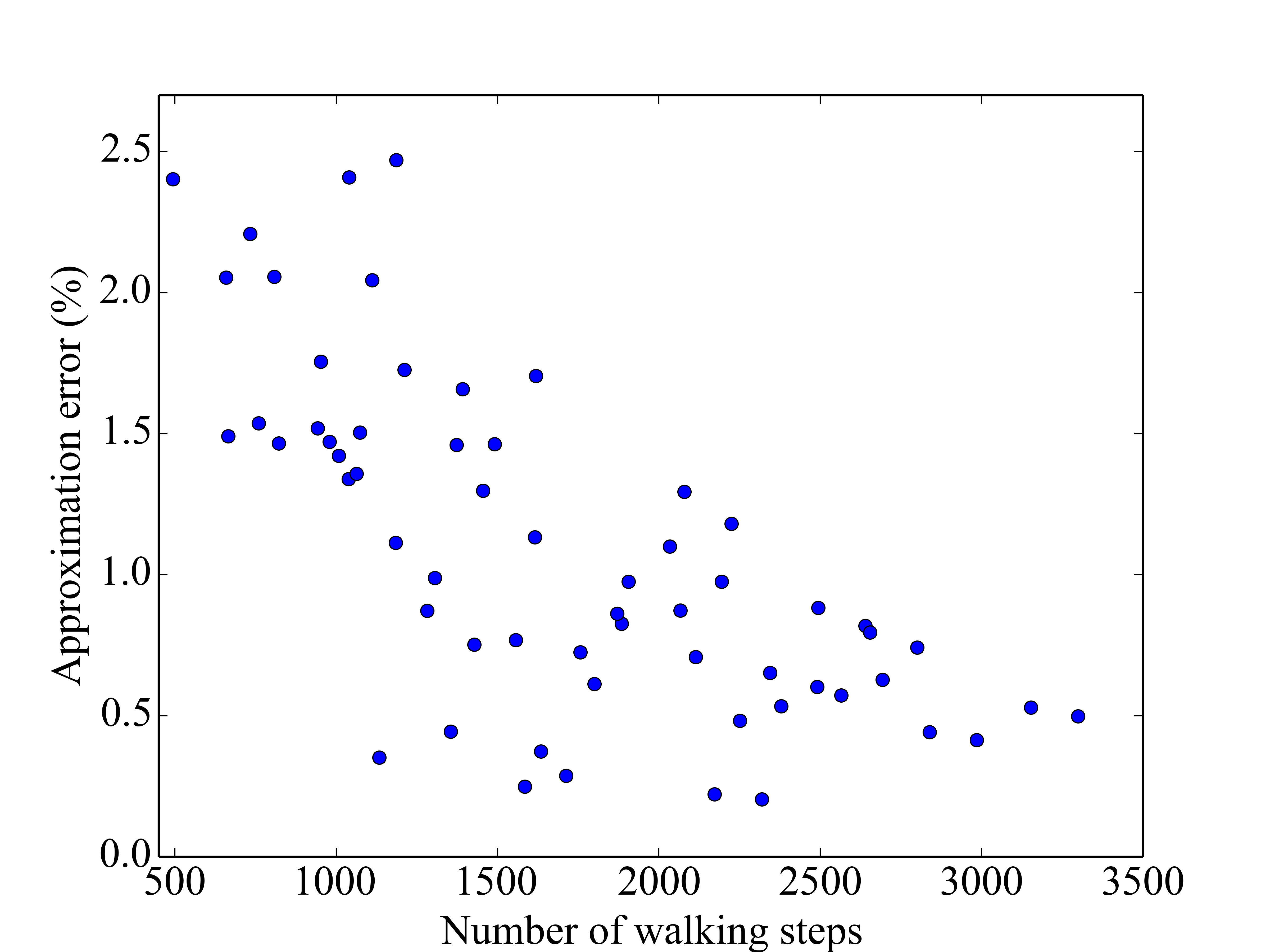}
\caption{Approximation error with respect to the number of walking steps.}
\label{fig:walking}
\end{figure} 

\begin{figure*}
\centering
\includegraphics[trim = 0mm 37mm 0mm 40mm ,clip,
width=500pt,height=150pt]{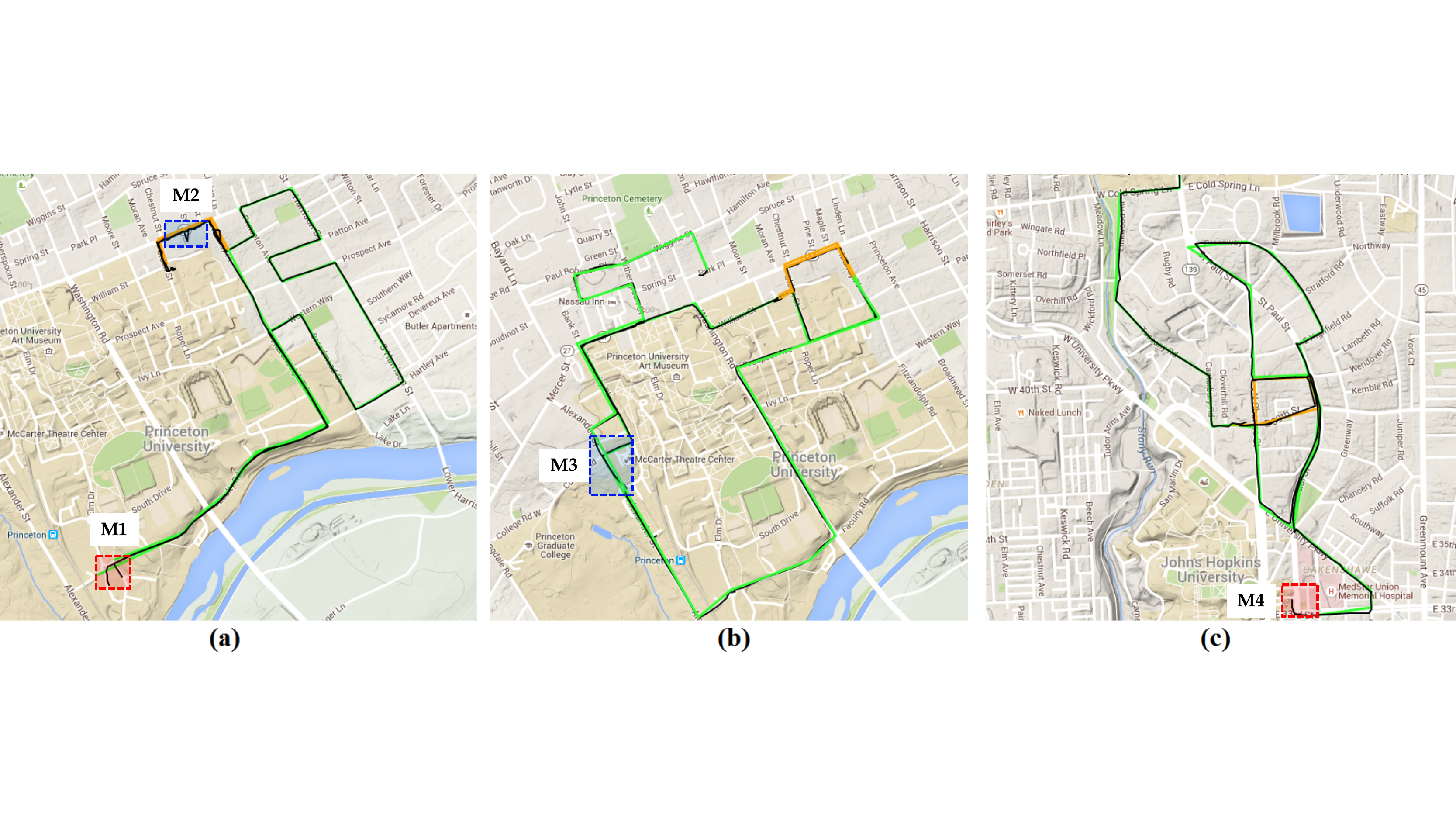}
\caption{Trajectories of three different users. Starting from the left
and moving to right: (a) the first user was located in Princeton and
carried a Galaxy S4 i9500, (b) the second user was located in
Princeton and carried an iPhone 6, and (c) the third user was
located in Baltimore and carried an iPhone 6S. The green and orange lines demonstrate the estimated user's paths during driving and walking, respectively. The black line is the actual user's trajectory reported by GPS data.}
\label{fig:alltra}
\end{figure*}

\subsubsection{End-to-end evaluation}
In order to provide an end-to-end evaluation, we evaluated the accuracy of PinMe using Dataset \#2. As discussed in Section \ref{ALGORITHMS}, we have implemented two different versions of \textit{walkingUserTracker}. For this evaluation, we used the second version, which assumes that the user is within a small area around his vehicle after he leaves the vehicle. Fig.~\ref{fig:alltra} demonstrates the actual trajectories of the users' movements (as provided by GPS sensor) along with 
the estimated trajectories (as provided by PinMe). As illustrated in this 
figure, for all three data streams, which were collected by three different 
users while carrying three different smartphones, the actual trajectories 
of the users' movements were very similar to the estimated ones provided 
by PinMe. However, we observed four mismatch areas (bounded by red/blue boxes 
in Fig.~\ref{fig:alltra}). In the first and last areas ($M1$ and $M4$), the 
starting point of the actual driving path was slightly different from the 
point discovered by PinMe due to the similarities between two nearby 
intersections marked on the map. In two other mismatch areas, PinMe more 
accurately located the user than GPS. The GPS trajectory shows that the user's 
vehicle was off the road ($M2$). Furthermore, it indicates that the user was 
off the sidewalk when he was walking ($M3$). In these two cases, we checked the validity of PinMe's trajectories with the users, and they confirmed that the results provided by PinMe show the actual trajectory in $M2$ and $M3$. 

Based on our experimental results, we can say that the location estimation 
accuracy of \textit{carTracker} was independent of the user's smartphone and 
vehicle. This was expected for two reasons. First, PinMe utilizes sensory data, 
which do not correlate with the smartphone model (air pressure, heading, and 
acceleration), as opposed to PowerSpy \cite{BON} that uses power 
consumption, which highly correlates with the smartphone model. Second, as 
described in Section \ref{PMECH}, \textit{carTracker} mainly relies on air 
pressure and heading to track the vehicle when the user is driving -- these 
data are not correlated with the vehicle model, as opposed to acceleration 
data that are correlated with the vehicle model due to the existence of 
vibrations caused by the engine of the running vehicle \cite{ACCO}.

\section{Countermeasures}
\label{CONT}
In this section, we briefly describe several countermeasures (along with their shortcomings) for mitigating the risks of attacks against location privacy.

\subsection{Adaptive sampling rate}
\label{ADAPS}
Limiting the sampling rate of sensors can potentially limit the amount
of information leaked by the smartphone. In order to briefly discuss how
the accuracy of PinMe might be negatively impacted if the sampling rate 
decreases, we examined \textit{carTracker} using sensory data collected
at different sampling rates. Fig. \ref{fig:sampling} shows how the
average approximation error of \textit{carTracker} changes with respect
to the sampling rate. As we decrease the sampling rate, the
approximation error only slightly increases for this algorithm (even when the 
sampling rate is around $0.1Hz$). However, based on our empirical results, the 
accuracy of \textit{carTracker} suddenly drops when the sampling rate becomes 
very low (i.e., below $0.02Hz$) since the algorithm cannot detect the 
intersection (when the car turns) anymore. Many benign applications (for 
example, fitness tracker \cite{FITNESS_HEALTH_1} and  fall detection 
\cite{FALL_DETECT}) require a sampling frequency larger than $0.1Hz$, and 
thus decreasing the sampling rate of sensors below $0.1Hz$, to prevent 
a PinMe attack, would reduce the efficiency, efficacy, and utility of trusted 
applications as well. 

\begin{figure}
\centering
\includegraphics[trim = 5mm 5mm 5mm 5mm ,clip,
width=200pt,height=150pt]{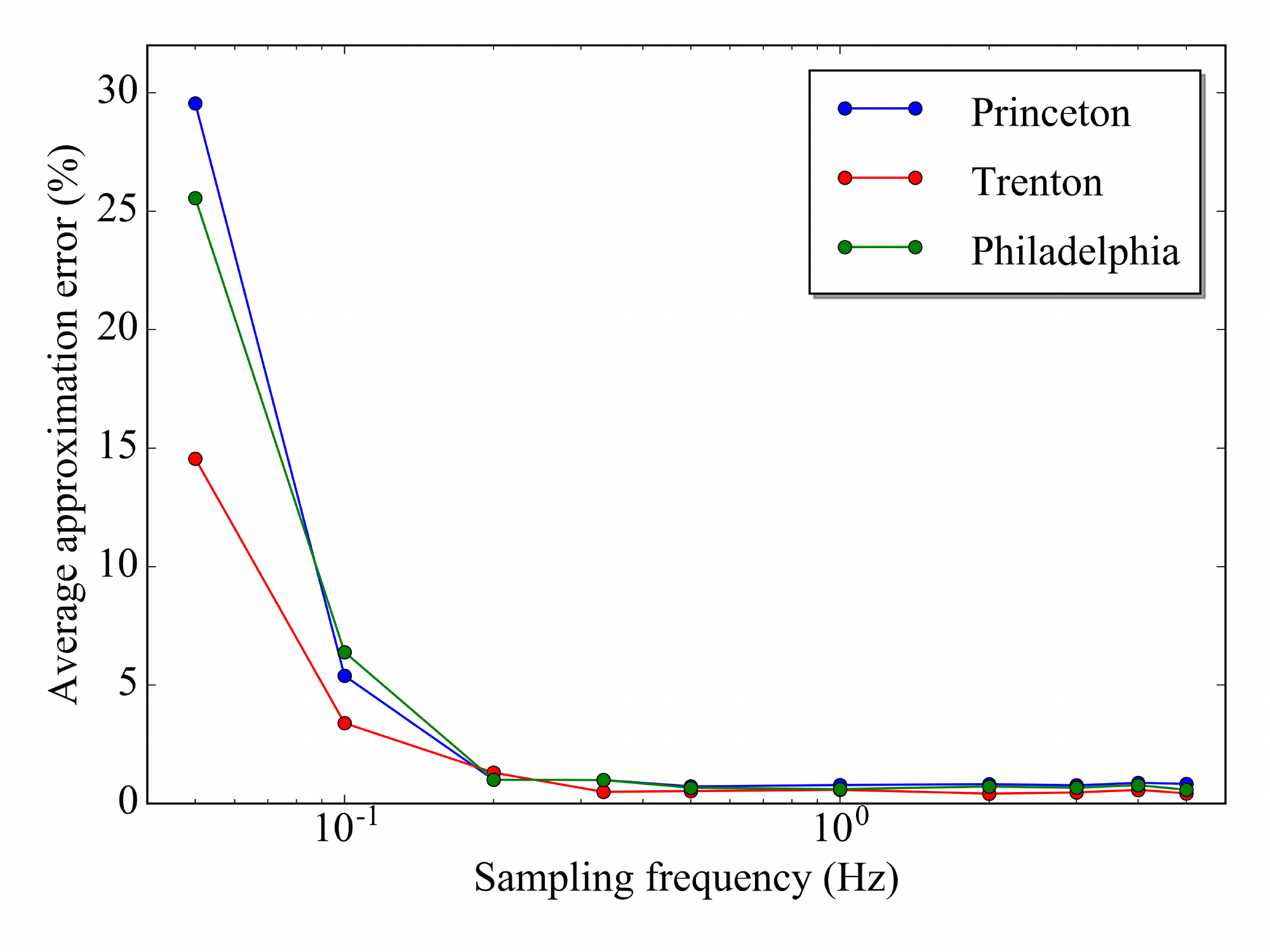}
\caption{Average approximation error of \textit{carTracker} with respect to the sampling frequency.}
\label{fig:sampling}
\end{figure}

Utilizing context-aware sampling mechanisms, which can adaptively control 
sensor sampling rates, may be an alternative approach to maximizing utility 
and minimizing information leakage. For example, consider a mechanism that 
changes the maximum allowable sampling rate of the sensors based on user's 
current activity. Such a mechanism can allow a fitness tracking application 
to obtain very frequent samples from the accelerometer when the user is 
running and only allow infrequent sampling when the user is driving.

\subsection{Risk-evaluation mechanism} 
\label{RISK_EVAL}
Generally, a risk-evaluation mechanism aims to share the smartphone's data in such a way that certain kinds of inferences cannot be drawn. It examines if a set of sensory/non-sensory data collected by an application can leak sensitive information about the user, and blocks an application upon the detection of a potential information leakage. A few recent research efforts have been geared towards risk-evaluation mechanisms that can be implemented on the 
smartphone to ensure user privacy \cite{REM1,IP,LEAVE_ME}. For instance, Chakraborty et al.~\cite{IP} have proposed ipShield, a framework to control the sensory data that are accessible by various applications installed on a smartphone. Their risk-evaluation mechanism continuously examines what inferences can be made from the shared sensory/non-sensory information. 
  
Zhang et al.~\cite{LEAVE_ME} proposed a defense against runtime-information-gathering attacks in which a malicious app runs side-by-side with a target application (a victim) and performs runtime information gathering (RIG). They suggested temporarily stopping the applications that are potentially able to collect data from a sensitive application or killing applications that may be collecting side-channel information in the background while the foreground application performs sensitive tasks. They discuss two suspicious activities that can reveal maliciousness of an application: (1) a high sampling rate needed for continuous monitoring, (2) the presence of a correlation between an application's activity and the activity of a sensitive application. The location estimation algorithms described in our paper need a much lower sampling frequency (for example, as shown earlier, $0.1Hz$ led to accurate results for \textit{carTracker}) than the frequency used in many previous attacks (for example, ACComplice \cite{ACCO} uses a sampling rate of $30Hz$). Therefore, sampling rate cannot be solely used to reveal the malicious activity of PinMe. Furthermore, PinMe does not require any data from other applications since it directly collects permission-free data, therefore, there is no correlation between its activity and other applications' activities. Finally, their defense relies on monitoring application-specific files, which are no longer accessible in Android M \cite{SIMON}. Thus, the approach discussed in \cite{LEAVE_ME} does not address PinMe.

\subsection{Sensor data manipulation}
Sensor data manipulation enables the user to manipulate or add noise to the 
content of collected sensory data when he is apprehensive about
sensor data abuse in certain sensing applications. Typical data manipulation 
approaches include rounding the values in the sensory data to approximate 
values, replacing particular sensor readings by previously-recorded readings, 
and adding random noise to the sensory data. However, as mentioned
earlier, unlike many previous attacks, PinMe relies on several
macro-level features extracted from sensory data. As a result, it is
robust against several potential sources of noise. For example, for
\textit{planeTracker} described in Section \ref{ALGORITHMS}, it only
extracts the aviation phases of the plane from noisy acceleration
readings (as apposed to the actual displacement) from which it estimates
the flight duration. As shown in Fig.~\ref{fig:both}
(Fig.~\ref{fig:destination}), \textit{planeTracker} was able to find a
set of three routes (airports) that includes the actual flight route
(destination airport), with a high level of accuracy, even when the
approximated duration and elevation are assumed to be inaccurate due to
the presence of noise (up to $10\%$ for flight duration and $5m$ for 
elevation). 

Adding significant noise to sensory readings or replacing data with 
previously-recorded data may significantly reduce the utility of trusted 
applications relying on such sensory data. 

\subsection{Turn-off switch}
A hardware turn-off switch that lets the user quickly and easily turn off all 
sensors or a sensor-free mode implemented in the operating system in which no 
application can obtain sensory information enables the user to easily stop 
information leakage when he suspects that there might be privacy 
risks. For example, the user can turn off all sensors when he is driving to 
ensure that no application can track him.

\section{Related Work}
\label{RELATED}
\begin{table*}[t] 
\caption{Comparison of different user-tracking mechanisms} 
\centering 
\begin{tabular}{l|c|c|c|c|c|c} 
\hline\hline 
Tracking mechanism & \#Activity &  Prior info. & Training & OS & Sampling freq. & Device/Vehicle dependence \\ [0.5ex]
\hline 
PowerSpy \cite{BON} & 1 &  $Y$ & $Y$ & Andorid & N/A & $Y$\\ [0.5ex]
\hline 
ACComplice \cite{ACCO} & 1 &  $Y$ & $Y$ & Android and iOS & $30 Hz$ & $Y$\\ [0.5ex]
\hline 
Tracking Metro \cite{SUB} & 1 &  $Y$  & $Y$ & Android and iOS & $10 Hz$ & $N/A$ \\ [0.5ex]
\hline 
From Pressure\cite{BARO} & 1 &  $Y$  & $N$ & N/A & $30 Hz$ & $N$ \\ [0.5ex]
\hline 
Narain et al. \cite{NARAIN} & 1 &  $N$  & $N$ & Android & 20-100 & $Y$ \\ [0.5ex]

\hline 
RIG \cite{NAVEED} & 1 &  $Y$  & $N$ & Android & $50 Hz$ & $N$ \\ [0.5ex]

\hline 
PinMe & 4 & $N$ & $N$ & Android and iOS & $5 Hz$ ($0.1Hz$ for driving) & $N$\\ [0.5ex]

\hline %inserts single line 
\end{tabular}
\label{table:COMP}
\end{table*}

Several prior research studies have demonstrated the use of smartphone sensors in diverse application domains. The use of accelerometer for activity monitoring has been widely discussed in the literature \cite{ACCEL1,ACCEL2,ACCEL3,ACCEL4}. Furthermore, recent research articles have discussed the feasibility of using air pressure measurements for indoor positioning \cite{FLOORD1,FLOORD2}, in 
particular, floor detection. 

Moreover, as briefly mentioned in 
Section \ref{INT}, a few recent research efforts have demonstrated the 
feasibility of obtaining valuable information about the smartphone's location without accessing the GPS. In the following, we discuss them in more detail.

PowerSpy \cite{BON} demonstrated that an adversary can estimate the user's location by processing the power consumption information of the device when 
he is driving through a known set of routes. As mentioned in \cite{BON},
this user location mechanism has the following limitations: (i) it requires a 
massive training dataset of power profiles associated with GPS coordinates, 
(ii) since the power profiles of different smartphones vary significantly from 
each other, in order to construct the training dataset, the attacker 
needs to measure the power consumption of many devices while driving, 
(iii) it assumes that there is enough variability in the device's power 
consumption along a route such that it exhibits unique features, (iv) it is 
only applicable to Android devices, and (v) it is able to detect the complete 
driving path in only $45$\% of the trials in the in the best-case scenario using
HTC Desire for data collection and a small set of possible routes (the 
estimation accuracy significantly worsened when other smartphones were used in 
the experiment).

ACComplice \cite{ACCO} showed that continuous measurements of acceleration in 
smartphones can reveal user location while driving. It has four main 
limitations: (i) it requires a training dataset that contains data on 
multiple car trips through each potential traveling route, with the smartphone 
constantly collecting motion sensor data, (ii) since it mainly relies on 
smartphone's acceleration data, the noise in sensor readings, e.g, due to 
different road conditions, can significantly affect its accuracy, (iii) it 
returns several (usually more than 10) potential driving paths, and (iv) 
device acceleration needs to be measured at a relatively high frequency ($30 Hz$). We attempted to implement ACComplice\cite{ACCO}. However, we observed that 
the accelerometer measurements alone were \textit{extremely noisy} and 
led to results that were much weaker than those reported in \cite{ACCO}. 
Hence, for a fair comparison, we simply use the accuracy reported in 
that paper in our comparison. In \cite{ACCO}, ACComplice is evaluated using 
only two driving paths. When the initial point was not given to the algorithm, 
for each test, it was able to return two clusters of possible starting points (each including five points) such that the starting point was within one of the clusters. Knowing the initial location, it could only partially find the driving paths (it correctly found 18 out of 23 routes for one test case 
and 9 out of 12 routes for the other).  

Narain et al. \cite{NARAIN} demonstrated that an Android app can infer traveled routes, without the users' knowledge, using gyroscope, accelerometer, and magnetometer readings. In their attack, gyroscope readings have been used as the main source of data. Further, accelerometer and magnetometer readings have been utilized to reduce noise and refine the results. Their proposed approach has four main limitations: (i) the attack returns 10 potential driving paths, (ii) in real-world experiments, they report a probability of only 30\% of inferring a list of 10 routes containing the true route, (iii) the proposed algorithm only works for driving, and (iv) since this attack requires a very high sampling rate ($20-100Hz$), their application can be easily marked as ``malicious'' using the approach described in \cite{LEAVE_ME}.

Zhou et al. \cite{NAVEED} discussed an attack based on acoustic
information leakage from another application. Their approach processes
the sequence of acoustic data generated by the smartphone's speaker when
the user is driving and using a navigational application. This attack
does not depend on the vehicle and device, and constructing real-world
attack-specific dataset (it constructs a dataset based on simulations). 
However, they assume that attacker knows the user's start location or a place 
on his route and the rough area he goes (e.g., city) to find some points of 
interest and implicitly ignore the possible loops/rerouting. Furthermore, if 
the user goes to unlabeled places that is not likely to be included in the 
constructed dataset of points of interest (for example, if he parks his 
vehicle far from a point of interest), the approach discussed in 
\cite{NAVEED} is unable to return the user's trajectory. In addition, using the approach described in \cite{LEAVE_ME}, this attack can be detected due to its high high sampling rate ($50Hz$). Finally, if the user simply turns off the speech guidance mechanism of the navigation application, this attack is not applicable anymore. 

Ho et al. \cite{BARO} presented an approach that uses dynamic time
warping (DTW) algorithms (i.e., a time-series alignment algorithm in
which two signals are compared against each other by means of a cost
matrix) to track a vehicle using air pressure readings sampled at
$30Hz$. DTW is used to compare the sequence of air pressure data samples
with that of different candidate paths. However, in real-world
scenarios, unfortunately, the search space of all candidate paths can be
very large. If path loops are included, the search space may be infinite. 
They assumed that the path does not contain any loop and examined two 
DTW-based methods. For the first one, the median error is reported 
to be around $800m$ (when median error for a random walk was only $1600 m$). 
Considering prior knowledge about the user to limit the area of interest and 
reduce search complexity, the second algorithm offers a median error of 
$60 m$. However, as mentioned in \cite{BARO}, the second approach does not 
scale well for large maps.

Hua et al. \cite{SUB} demonstrated that acceleration data can provide valuable location-related information when the user is traveling on a train.  As mentioned in \cite{SUB}, the tracking method has two main limitations: (i) similar to 
the above-mentioned methods, it requires a large training 
dataset collected by the attacker while traveling through different potential 
paths, and (ii) it is difficult to provide a high level of location estimation 
accuracy due to various types of noise in the training data.

Table \ref{table:COMP} compares different location mechanisms and highlights 
the advantages of PinMe. Our experimental results indicate that, without knowing the initial location, 
PinMe was able to return a single accurate driving path that is very similar to the trajectory provided by GPS readings. We believe that PinMe 
is able to return very accurate results since it mainly relies on noise-robust 
features extracted from barometer and magnetometer measurements. Moreover, 
unlike previously-proposed mechanisms, PinMe does not require measurements 
on a set of possible routes in advance. Therefore, our proposed attack is 
also more scalable. Unlike PinMe, the above-mentioned attacks only 
estimate the user's location during a single activity. Moreover, they commonly assume that the adversary has substantial prior information about the user's initial location. This knowledge is required because the attacker needs to collect a set of sensory data for different potential routes in 
advance and construct an attack-specific training database (e.g., in 
\cite{BON}) or the location estimation algorithm does not scale well for
a large area of interest (e.g., in \cite{BARO}). 

\section{Discussion}
\label{DISC}
In this section, we discuss three items not yet explained in detail. First, we discuss limitations of the proposed mechanism. Second, we describe how we took advantage of the interdependence between activities in our algorithms. We then discuss how PinMe can 
also be used as a stand-alone location mechanism, and how it can be used to enhance the security of autonomous vehicles. 

\subsection{Limitations}
\label{LIMITATIONS}
Next, we briefly discuss four potential limitations of PinMe. 

PinMe uses the history of smartphone IP addresses to infer the last city in 
which the user was connected to a WiFi network. In fact, it assumes that the 
user is directly connected to the Internet. Thus, if the user utilizes an 
anonymous communication service, e.g., Tor \cite{TOR}, PinMe may fail to 
locate the user. However, as mentioned later, the interdependence between activities can be used to resolve this limitation. 

Moreover, PinMe relies significantly on the variability
of elevations and route directions. Therefore, PinMe might be unable
to estimate the user's location if the user only moves in grid routes, e.g., 
some parts of Manhattan, NY, in which the roads are almost flat and parallel to 
each other. Furthermore, since PinMe relies on publicly-available datasets, 
the existence of erroneous data in auxiliary datasets given to PinMe may 
reduce the location estimation accuracy. For example, OSM navigational maps 
do not typically include very recent constructions/detours. Therefore, if 
the user travels through a new road that has not been added to the map, 
PinMe may fail to track the user.

Despite the above-mentioned limitations, PinMe presents a significant advance 
in state-of-the-art smartphone-based user location, since it enables an 
attacker to scale up the attack against location privacy by minimizing attack 
requirements and offers a high location estimation accuracy.

\subsection{Interdependence of activities}
\label{INTER}
As described earlier in Section \ref{ALGORITHMS}, we designed four different independent algorithms for tracking the user during four different activities. Although the user's activities may seem independent of each other at first 
glance, there exists an interdependence between them due to physical 
constraints imposed by the world and the user's movement. 

In particular, we make two observations. First, the users always walk between 
other activities (driving, traveling on a train, and traveling on a plane), 
and therefore, certain sequences of activities are not feasible. For example, 
the user cannot get on a plane as soon as he stops driving. This helps our tailored classifier algorithm to remove impossible cases. 

Second, the final location of the user after performing each activity \textit{roughly} determines the initial location of the next activity. However, since the precision of the estimated location determined by different algorithms might differ from each other, combining the results from different algorithms to get an accurate trajectory is not usually straightforward. For example, consider the 
following scenario: a user takes a flight that lands at airport A, then walks 
for a few hundred meters to reach his car, and eventually drives to his home 
from the airport. In order to track the user, PinMe utilizes 
\textit{flightTracker}, \textit{walkingUserTracker}, and \textit{carTracker}, 
respectively.  \textit{flightTracker} returns departure and destination 
airports, whereas \textit{carTracker} and \textit{walkingUserTracker} 
return a trajectory with an accuracy comparable to GPS. If PinMe relies on the assumption that the initial location for each activity is \textit{accurately} determined by the previous activity, then it fails to provide an accurate estimation of the user's trajectory in the above-mentioned scenario since the location returned by the first activity provides an inaccurate initial point for \textit{carTracker} (the whole airport area is marked as a single point with fixed GPS coordinates on navigational maps). However, the interdependence between activities still provides valuable pieces of information in this scenario. First, \textit{flightTracker} returns the destination airport from which the current city can be identified even if the user has not connected to any WiFi network yet or 
is using an anonymous communication service, e.g., Tor \cite{TOR}. Second, the final location of the user after performing each activity can significantly bound our area of interest. This has been used in our end-to-end evaluation, where the \textit{walkingUserTracker} algorithm assumes that the user's initial location, when he starts walking, is within a small area around the final location of the user estimated by \textit{carTracker}. 

\subsection{PinMe as an alternative to GPS}
Next, we first describe drawbacks of traditional GPS systems. We then describe 
why PinMe can offer a more secure navigation mechanism for autonomous vehicles.

With the widespread use of GPS receivers in modern vehicles, ranging from 
yachts to autonomous cars, the security of GPS has garnered ever-increasing 
attention in recent years. GPS receivers compare timestamped signals from a 
constellation of satellites, inferring their position through computations
on the lightspeed lag from each signal. Several research studies 
\cite{GPSS1,GPSS2,GPSS3} have demonstrated the feasibility of faking the 
satellite signals needed for positioning and mentioned that security attacks 
against the GPS signals used in autonomous vehicles may lead to disastrous 
consequences. 

Unfortunately, protecting GPS signals against spoofing is difficult for three 
reasons. First, the computational load associated with cryptographic 
signatures on the signal is high. Second, it is impossible to use a 
challenge-response protocol since the communication channel between the 
satellites and GPS receiver is unidirectional, i.e., the receiver cannot 
transmit data to the satellites. Third, the implementation of new 
algorithms/mechanisms, which need modifications to the GPS infrastructure, 
is difficult and costly.

As demonstrated in Section \ref{EVAL}, PinMe was able to accurately 
(comparable to GPS) locate the user during different activities. A 
slightly modified version of PinMe can be implemented on autonomous vehicles, 
e.g., driverless cars, as a stand-alone in-vehicle positioning system. For 
example, air pressure and heading sensors can be added to driverless vehicles,
enabling  sensory data to be processed by on-vehicle processing units. Odometer 
readings are easily accessible to in-vehicle processing units and can be 
used to further improve the accuracy of PinMe. Since PinMe does not collect 
sensory data from any remote sources, it is resilient against remote attacks, 
assuming that navigational/elevation maps provided by Google \cite{GAPI} and 
weather reports given by The Weather Channel \cite{WET} are accurate. 

\section{Conclusion} 
\label{CONC}
This paper highlighted the unintended consequences of letting third-party 
applications access smartphone's presumably non-critical data. We proposed an 
attack on location privacy in which the attacker (i) needs no prior knowledge 
of the area of interest, (ii) does not need to construct an
attack-specific training dataset, and (iii) does not collect data at a high sampling rate. 

We  demonstrated that there is no need to construct an attack-specific 
dataset to compromise location privacy. Evaluation of the proposed
user-location mechanism demonstrated that it is feasible to gain 
sensitive information about the user's location without accessing location 
services, e.g., GPS. It suggests that the threat of unintended information 
leakage on the location of smartphone owners is far beyond what is
currently thought possible. Indeed, even seemingly benign sensory/non-sensory 
data gathered by a smartphone can leak critical information about the user.
Therefore, they should be proactively protected from third-party applications.
\bibliographystyle{IEEEtran}
\bibliography{Arxiv_Version}

\begin{thebibliography}{10}
\providecommand{\url}[1]{#1}
\csname url@samestyle\endcsname
\providecommand{\newblock}{\relax}
\providecommand{\bibinfo}[2]{#2}
\providecommand{\BIBentrySTDinterwordspacing}{\spaceskip=0pt\relax}
\providecommand{\BIBentryALTinterwordstretchfactor}{4}
\providecommand{\BIBentryALTinterwordspacing}{\spaceskip=\fontdimen2\font plus
\BIBentryALTinterwordstretchfactor\fontdimen3\font minus
  \fontdimen4\font\relax}
\providecommand{\BIBforeignlanguage}[2]{{%
\expandafter\ifx\csname l@#1\endcsname\relax
\typeout{** WARNING: IEEEtran.bst: No hyphenation pattern has been}%
\typeout{** loaded for the language `#1'. Using the pattern for}%
\typeout{** the default language instead.}%
\else
\language=\csname l@#1\endcsname
\fi
#2}}
\providecommand{\BIBdecl}{\relax}
\BIBdecl

\bibitem{SPAM1}
R.~Shokri, G.~Theodorakopoulos, J.-Y. Le~Boudec, and J.-P. Hubaux,
  ``Quantifying location privacy,'' in \emph{Proc. IEEE Symp. Security and
  Privacy}, 2011, pp. 247--262.

\bibitem{SEC1}
E.~Owusu, J.~Han, S.~Das, A.~Perrig, and J.~Zhang, ``{ACCessory}: Password
  inference using accelerometers on smartphones,'' in \emph{Proc. ACM Wkshp.
  Mobile Computing Systems \& Applications}, 2012, p.~9.

\bibitem{SEC2}
L.~Cai and H.~Chen, ``{TouchLogger}: Inferring keystrokes on touch screen from
  smartphone motion,'' in \emph{Proc. HotSec}, 2011.

\bibitem{SEC3}
E.~Miluzzo, A.~Varshavsky, S.~Balakrishnan, and R.~R. Choudhury, ``Tapprints:
  Your finger taps have fingerprints,'' in \emph{Proc. ACM Int. Conf. Mobile
  Systems, Applications, and Services}, 2012, pp. 323--336.

\bibitem{SEC4}
Z.~Xu, K.~Bai, and S.~Zhu, ``{Taplogger}: Inferring user inputs on smartphone
  touchscreens using on-board motion sensors,'' in \emph{Proc. ACM Conf.
  Security and Privacy in Wireless and Mobile Networks}, 2012, pp. 113--124.

\bibitem{SEC5}
A.~J. Aviv, K.~Gibson, E.~Mossop, M.~Blaze, and J.~M. Smith, ``Smudge attacks
  on smartphone touch screens,'' in \emph{Proc. USENIX Wkshp. Offensive
  Technologies}, vol.~10, 2010, pp. 1--7.

\bibitem{SEC6}
Y.~Wang, K.~Streff, and S.~Raman, ``Smartphone security challenges,''
  \emph{Computer}, vol.~45, no.~12, pp. 52--58, 2012.

\bibitem{SEC7}
S.~Jana and V.~Shmatikov, ``Memento: Learning secrets from process
  footprints,'' in \emph{Proc. IEEE Symp. Security and Privacy}, 2012, pp.
  143--157.

\bibitem{NAVEED}
X.~Zhou, S.~Demetriou, D.~He, M.~Naveed, X.~Pan, X.~Wang, C.~A. Gunter, and
  K.~Nahrstedt, ``Identity, location, disease and more: Inferring your secrets
  from {Android} public resources,'' in \emph{Proc. ACM SIGSAC Conf. Computer
  \& Communications Security}, 2013, pp. 1017--1028.

\bibitem{LSEC1}
A.~Papliatseyeu and O.~Mayora, ``Mobile habits: Inferring and predicting user
  activities with a location-aware smartphone,'' in \emph{Proc. Symp.
  Ubiquitous Computing and Ambient Intelligence}, vol.~13, 2009, pp. 343--352.

\bibitem{LSEC2}
D.~He, ``Security threat to {Android} apps,'' Master's thesis, Dept. Computer
  Science, University of Illinois at Urbana-Champaign, 2014.

\bibitem{SPAM2}
R.~Shokri, P.~Papadimitratos, G.~Theodorakopoulos, and J.-P. Hubaux,
  ``Collaborative location privacy,'' in \emph{Proc. IEEE Int. Conf. Mobile
  Adhoc and Sensor Systems}, 2011, pp. 500--509.

\bibitem{ACT}
``{Geolocation Privacy},''
  \url{http://www.gps.gov/policy/legislation/gps-act/}, accessed: 2016-10-20.

\bibitem{LOC1}
K.~W.~Y. Au, Y.~F. Zhou, Z.~Huang, P.~Gill, and D.~Lie, ``Short paper: A look
  at smartphone permission models,'' in \emph{Proc. ACM Wkshp. Security and
  Privacy in Smartphones and Mobile Devices}, 2011, pp. 63--68.

\bibitem{LOC2}
A.~P. Felt, H.~J. Wang, A.~Moshchuk, S.~Hanna, and E.~Chin, ``Permission
  re-delegation: Attacks and defenses,'' in \emph{Proc. USENIX Security
  Symposium}, 2011.

\bibitem{ATT1}
T.~Watanabe, M.~Akiyama, and T.~Mori, ``{RouteDetector}: Sensor-based
  positioning system that exploits spatio-temporal regularity of human
  mobility,'' in \emph{Proc. USENIX Wkshp. Offensive Technologies}, 2015.

\bibitem{BON}
Y.~Michalevsky, A.~Schulman, G.~A. Veerapandian, D.~Boneh, and G.~Nakibly,
  ``Powerspy: Location tracking using mobile device power analysis,'' in
  \emph{Proc. USENIX Security Symposium}, 2015, pp. 785--800.

\bibitem{SUB}
J.~Hua, Z.~Shen, and S.~Zhong, ``We can track you if you take the metro:
  Tracking metro riders using accelerometers on smartphones,'' \emph{arXiv
  preprint arXiv:1505.05958}, 2015.

\bibitem{ACCO}
J.~Han, E.~Owusu, L.~T. Nguyen, A.~Perrig, and J.~Zhang, ``{ACComplice}:
  Location inference using accelerometers on smartphones,'' in \emph{Proc. IEEE
  Int. Conf. Communication Systems and Networks}, 2012, pp. 1--9.

\bibitem{BARO}
B.-J. Ho, P.~Martin, P.~Swaminathan, and M.~Srivastava, ``From pressure to
  path: Barometer-based vehicle tracking,'' in \emph{Proc. ACM Int. Conf.
  Embedded Systems for Energy-Efficient Built Environments}, 2015, pp. 65--74.

\bibitem{LEAVE_ME}
N.~Zhang, K.~Yuan, M.~Naveed, X.~Zhou, and X.~Wang, ``Leave me alone: App-level
  protection against runtime information gathering on {Android},'' in
  \emph{Proc. IEEE Symp. Security and Privacy}, 2015, pp. 915--930.

\bibitem{PHYDAM}
H.~Zhu, S.~Du, M.~Li, and Z.~Gao, ``Fairness-aware and privacy-preserving
  friend matching protocol in mobile social networks,'' \emph{IEEE Trans.
  Emerging Topics in Computing}, vol.~1, no.~1, pp. 192--200, 2013.

\bibitem{HOH}
B.~Hoh, M.~Gruteser, H.~Xiong, and A.~Alrabady, ``Enhancing security and
  privacy in traffic-monitoring systems,'' \emph{Pervasive Computing}, vol.~5,
  no.~4, pp. 38--46, 2006.

\bibitem{REPORT}
``{mHealth App Developer Economics 2014},''
  \url{http://mhealtheconomics.com/mhealth-developer-economics-report/},
  accessed: 2016-10-20.

\bibitem{OSM}
``Openstreetmap,'' \url{https://www.openstreetmap.org}, accessed: 2016-10-20.

\bibitem{GAPI}
``{Google Maps API},'' \url{https://developers.google.com/maps}, accessed:
  2016-10-20.

\bibitem{UAPI}
``{Maps, Imagery, and Publications},''
  \url{http://www.usgs.gov/pubprod/maps.html}, accessed: 2016-10-20.

\bibitem{WET}
``The weather channel,'' \url{https://weather.com/}, accessed: 2016-10-20.

\bibitem{OFLIGHT}
``{Airport and Route Data},'' \url{http://openflights.org/data.html}, accessed:
  2016-10-20.

\bibitem{GOOGLEMAP}
``Google maps,'' \url{https://www.google.com/maps}, accessed: 2016-10-20.

\bibitem{LF1}
M.~Balakrishnan, I.~Mohomed, and V.~Ramasubramanian, ``Where's that phone?:
  Geolocating {IP} addresses on {3G} networks,'' in \emph{Proc. ACM SIGCOMM
  Internet Measurement Conference}, 2009, pp. 294--300.

\bibitem{LF2}
S.~Triukose, S.~Ardon, A.~Mahanti, and A.~Seth, ``Geolocating {IP} addresses in
  cellular data networks,'' in \emph{Proc. Passive and Active Measurement},
  2012, pp. 158--167.

\bibitem{ACTF1}
J.~P{\"a}rkk{\"a}, M.~Ermes, P.~Korpip{\"a}{\"a}, J.~M{\"a}ntyj{\"a}rvi,
  J.~Peltola, and I.~Korhonen, ``Activity classification using realistic data
  from wearable sensors,'' \emph{IEEE Trans. Information Technology in
  Biomedicine}, vol.~10, no.~1, pp. 119--128, 2006.

\bibitem{ACTF2}
S.~J. Preece, J.~Y. Goulermas, L.~P. Kenney, and D.~Howard, ``A comparison of
  feature extraction methods for the classification of dynamic activities from
  accelerometer data,'' \emph{IEEE Trans. Biomedical Engineering}, vol.~56,
  no.~3, pp. 871--879, 2009.

\bibitem{ACTF3}
A.~M. Khan, Y.~Lee, S.~Lee, and T.-S. Kim, ``Human activity recognition via an
  accelerometer-enabled-smartphone using kernel discriminant analysis,'' in
  \emph{Proc. IEEE Int. Conf. Future Information Technology}, 2010, pp. 1--6.

\bibitem{ACTF4}
P.~Siirtola and J.~R{\"o}ning, ``Recognizing human activities
  user-independently on smartphones based on accelerometer data,'' \emph{Int.
  J. Interactive Multimedia and Artificial Intelligence}, vol.~1, no.~5, pp.
  39--45, 2012.

\bibitem{LSVM}
J.~A. Suykens and J.~Vandewalle, ``Least squares support vector machine
  classifiers,'' \emph{Neural Processing Letters}, vol.~9, no.~3, pp. 293--300,
  1999.

\bibitem{S_FORMULA}
``Atmospheric pressure at different altitudes,''
  \url{https://www.avs.org/AVS/files/c7/c7edaedb-95b2-438f-adfb-36de54f87b9e.pdf},
  accessed: 2016-10-20.

\bibitem{FORMULA}
D.~Jacob, \emph{Introduction to Atmospheric Chemistry}.\hskip 1em plus 0.5em
  minus 0.4em\relax Princeton University Press, 1999.

\bibitem{SENSORLOG}
``Sensorlog,'' \url{https://itunes.apple.com/us/app/sensorlog}, accessed:
  2016-10-20.

\bibitem{MONTE}
C.~Z. Mooney, \emph{Monte Carlo Simulation}.\hskip 1em plus 0.5em minus
  0.4em\relax Sage Publications, 1997, vol. 116.

\bibitem{FITNESS_HEALTH_1}
T.~M. Do, S.~W. Loke, and F.~Liu, ``{Healthylife}: An activity recognition
  system with smartphone using logic-based stream reasoning,'' in \emph{Proc.
  Int. Conf. Mobile and Ubiquitous Systems: Computing, Networking, and
  Services}, 2012, pp. 188--199.

\bibitem{FALL_DETECT}
P.~Kostopoulos, A.~I. Kyritsis, M.~Deriaz, and D.~Konstantas, ``{F2D:} a
  location aware fall detection system tested with real data from daily life of
  elderly people,'' \emph{Procedia Computer Science}, vol.~98, pp. 212--219,
  2016.

\bibitem{REM1}
J.~Tan, U.~Drolia, R.~Martins, R.~Gandhi, and P.~Narasimhan, ``{STOVEPipe}:
  Observable access control of user data for untrusted applications on mobile
  devices,'' in \emph{Proc. IEEE Int. Conf. Cloud Computing Technology and
  Science}, 2014, pp. 680--683.

\bibitem{IP}
S.~Chakraborty, C.~Shen, K.~R. Raghavan, Y.~Shoukry, M.~Millar, and
  M.~Srivastava, ``{ipShield}: A framework for enforcing context-aware
  privacy,'' in \emph{Proc. USENIX Symp. Networked Systems Design and
  Implementation}, 2014, pp. 143--156.

\bibitem{SIMON}
L.~Simon, W.~Xu, and R.~Anderson, ``Don't interrupt me while {I} type:
  Inferring text entered through gesture typing on {Android} keyboards,''
  \emph{Proc. Privacy Enhancing Technologies Symposium}, vol. 2016, no.~3, pp.
  136--154, 2016.

\bibitem{NARAIN}
S.~Narain, T.~D. Vo-Huu, K.~Block, and G.~Noubir, ``Inferring user routes and
  locations using zero-permission mobile sensors,'' in \emph{Proc. 2016 IEEE
  Symp. Security and Privacy}, 2016, pp. 397--413.

\bibitem{ACCEL1}
O.~D. Lara and M.~A. Labrador, ``A survey on human activity recognition using
  wearable sensors,'' \emph{IEEE Communications Surveys and Tutorials},
  vol.~15, no.~3, pp. 1192--1209, 2013.

\bibitem{ACCEL2}
X.~Su, H.~Tong, and P.~Ji, ``Activity recognition with smartphone sensors,''
  \emph{Tsinghua Science and Technology}, vol.~19, no.~3, pp. 235--249, 2014.

\bibitem{ACCEL3}
N.~D. Lane, E.~Miluzzo, H.~Lu, D.~Peebles, T.~Choudhury, and A.~T. Campbell,
  ``A survey of mobile phone sensing,'' \emph{IEEE Communications Magazine},
  vol.~48, no.~9, pp. 140--150, 2010.

\bibitem{ACCEL4}
S.~Hemminki, P.~Nurmi, and S.~Tarkoma, ``Accelerometer-based transportation
  mode detection on smartphones,'' in \emph{Proc. ACM Conf. Embedded Networked
  Sensor Systems}, 2013, p.~13.

\bibitem{FLOORD1}
H.~Xia, X.~Wang, Y.~Qiao, J.~Jian, and Y.~Chang, ``Using multiple barometers to
  detect the floor location of smart phones with built-in barometric sensors
  for indoor positioning,'' \emph{IEEE Sensors}, vol.~15, no.~4, pp.
  7857--7877, 2015.

\bibitem{FLOORD2}
B.~Li, B.~Harvey, and T.~Gallagher, ``Using barometers to determine the height
  for indoor positioning,'' in \emph{Proc. IEEE Int. Conf. Indoor Positioning
  and Indoor Navigation}, 2013, pp. 1--7.

\bibitem{TOR}
R.~Dingledine, N.~Mathewson, and P.~Syverson, ``Tor: The second-generation
  onion router,'' in \emph{Proc. USENIX Security Symp.}, 2004.

\bibitem{GPSS1}
A.~J. Kerns, D.~P. Shepard, J.~A. Bhatti, and T.~E. Humphreys, ``Unmanned
  aircraft capture and control via {GPS} spoofing,'' \emph{J. Field Robotics},
  vol.~31, no.~4, pp. 617--636, 2014.

\bibitem{GPSS2}
J.~S. Warner and R.~G. Johnston, ``{GPS} spoofing countermeasures,''
  \emph{Homeland Security Journal}, vol.~25, no.~2, pp. 19--27, 2003.

\bibitem{GPSS3}
S.~M. Giray, ``Anatomy of unmanned aerial vehicle hijacking with signal
  spoofing,'' in \emph{Proc. IEEE Int. Conf. Recent Advances in Space
  Technologies}, 2013, pp. 795--800.

\end{thebibliography}
\vspace{-1cm}
\begin{IEEEbiography}[
{\includegraphics[width=1.1in,height=1.25in,clip]{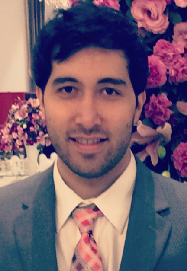}}]
{Arsalan Mosenia} is currently a postdoctoral research associate, jointly working with Profs. Mung Chiang (Purdue University) and Prateek Mittal (Princeton University). He received the B.Sc. degree in Computer Engineering from Sharif University of Technology in 2012, and the M.A. and Ph.D. in Electrical Engineering from Princeton University, in 2014 and 2016, respectively, under the supervision of Prof. Niraj K. Jha. 

He is broadly interested in investigating and addressing emerging security and privacy challenges in IoT. His research lies at the intersection of Internet of Things (IoT) and cyber-physical systems, machine learning, and information security. His work has uncovered fundamental security/privacy flaws in the design of multiple widely-used Internet-connected systems. His research impact includes several publications that are among the most popular papers of top-tier IEEE Transactions, multiple prestigious awards (including Princeton X, Princeton Innovation Fund, French-American Doctoral Exchange Fellowship, and Princeton IP Accelerator Fund), and extensive press coverage. Furthermore, at OpenFog Consortium, he is actively collaborating with Security Work Group, where he defines domain-specific security standards for fog computing, and Testbed Work Group, where he designs, builds, and examines novel fog-inspired real-world systems.

\end{IEEEbiography}
\vspace{-1.40cm}
\begin{IEEEbiography}[
{\includegraphics[width=1.1in,height=1.25in,clip]{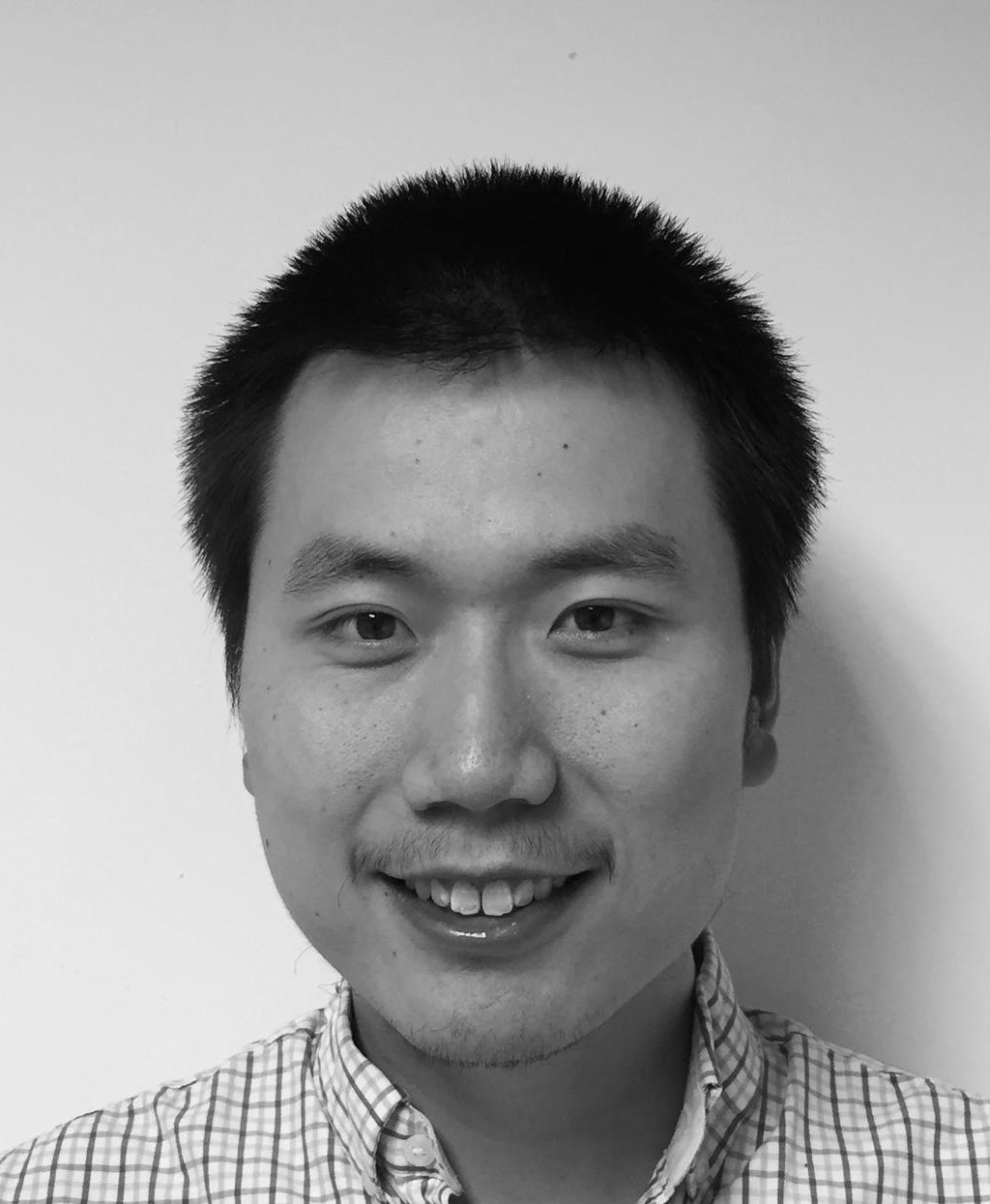}}]
{Xiaoliang Dai} received the B.Physics degree from Peking University, China, in 2014. He is currently a Ph.D. student in the Electrical Engineering Department at Princeton University. His research interests include machine learning for healthcare and security, Internet of Things, and novel mathematical models for TCAD simulations.
\end{IEEEbiography}
\vspace{-1cm}
\begin{IEEEbiography}[
{\includegraphics[width=1.1in,height=1.25in,clip]{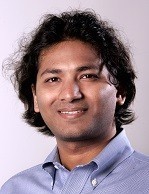}}]
{Prateek Mittal} is an assistant professor in the department of
Electrical Engineering at Princeton University. His research aims to
build secure and privacy-preserving communication systems. His
research interests include the domains of privacy enhancing
technologies, trustworthy social systems, and Internet/network
security.

His work has influenced the design of several widely used anonymity
systems, and is the recipient of several awards including an ACM CCS
outstanding paper. He served as the program co-chair for the FOCI and
the HotPETs workshops. He is the recipient of the NSF CAREER Award,
the Google Faculty Research Award, the M. E. Van Valkenburg research
Award, and Princeton Engineering Commendation List for Outstanding
Teaching.

Prior to joining Princeton University, he was a postdoctoral scholar
at University of California, Berkeley. He obtained his Ph.D. in
Electrical and Computer Engineering from University of Illinois at
Urbana-Champaign in 2012.
\end{IEEEbiography}

\newpage
\begin{IEEEbiography}[
{\includegraphics[width=1.1in,height=1.25in,clip]{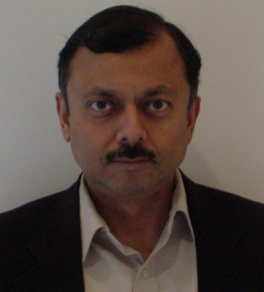}}]%
{Niraj K. Jha}(S'85-M'85-SM'93-F'98) received his B.Tech. degree in Electronics 
and Electrical Communication Engineering from Indian Institute of Technology, 
Kharagpur, India in 1981, M.S. degree in Electrical Engineering from S.U.N.Y. 
at Stony Brook, NY in 1982, and Ph.D. degree in Electrical Engineering from 
University of Illinois at Urbana-Champaign, IL in 1985. He is a Professor of 
Electrical Engineering at Princeton University. 

He is a Fellow of IEEE and ACM.  He received the Distinguished Alumnus Award 
from I.I.T., Kharagpur in 2014.  He is the recipient of the AT\&T Foundation 
Award and NEC Preceptorship Award for research excellence, NCR Award for 
teaching excellence, Princeton University Graduate Mentoring Award, and
six Outstanding Teaching Commendations from the School of Engineering
and Applied Sciences.

He has served as the Editor-in-Chief of IEEE Transactions on VLSI Systems and 
an Associate Editor of IEEE Transactions on Circuits and Systems I and II, 
IEEE Transactions on VLSI Systems, IEEE Transactions on Computer-Aided Design, 
IEEE Transactions on Computers, ournal of Electronic Testing: Theory and 
Applications, and Journal of Nanotechnology. He is currently serving as an 
Associate Editor of IEEE Transactions on Multi-Scale Computing
Systems and Journal of Low Power Electronics. He has served as 
the Program Chairman of the 1992 Workshop on Fault-Tolerant Parallel and 
Distributed Systems, the 2004 International Conference on Embedded and 
Ubiquitous Computing, and the 2010 International Conference on VLSI Design. 
He has served as the Director of the Center for Embedded System-on-a-chip 
Design funded by New Jersey Commission on Science and Technology.  He
has also served as an Associate Director of the Andlnger Center for
Energy and the Environment.

He has co-authored or co-edited five 
books that include two textbooks: Testing of Digital Systems (Cambridge 
University Press, 2003) and Switching and Finite Automata Theory, 3rd edition 
(Cambridge University Press, 2009).  He has co-authored 15 book 
chapters and more than 430 technical papers. He has coauthored 14 papers
that have won various awards, and another six that have received best
paper award nominations.  He has received 16 U.S. patents. He has served on the program committees of more than 150 conferences and workshops. His research 
interests include embedded computing, secure computing, machine learning,
smart healthcare, monolithic 3D IC design, low power hardware/software design, 
and computer-aided design of integrated circuits and systems. He has given 
several keynote speeches in the area of nanoelectronic design/test and
embedded systems.
\end{IEEEbiography}
\end{document}